\documentclass{jfm}

\usepackage{amsmath}
\usepackage{amssymb}
\usepackage{graphicx}
\usepackage{siunitx}
\usepackage[colorlinks]{hyperref}
\usepackage{cleveref}
\usepackage{placeins}
\usepackage{comment}
\usepackage[toc,page]{appendix}
\usepackage{mhchem}
\usepackage{enumerate}
\usepackage{tabularx}
\usepackage{multirow}
\usepackage{overpic}

\usepackage [english]{babel}
\usepackage [autostyle, english = american]{csquotes}
\MakeOuterQuote{"}

\usepackage{natbib}
\newcommand{\deriv}[2]{\frac{\partial #1}{\partial #2}}

\newcommand{\avg}[1]{\langle #1 \rangle}

\newcommand{\dH}{d_\mathrm{H}}
\newcommand{\ddH}{d/d_\mathrm{H}}
\newcommand{\dodH}{d_0/d_\mathrm{H}}
\newcommand{\Lint}{L_\mathrm{int}}

\let\olddelta\delta
\renewcommand{\delta}{{\olddelta}}
\let\oldDelta\Delta
\renewcommand{\Delta}{{\oldDelta}}

\newcommand{\nd}[1]{\tilde{#1}}
\newcommand{\ndf}{\tilde{f}}
\newcommand{\nddelta}{\tilde{\delta}}
\newcommand{\ndDelta}{\tilde{\Delta}}

\renewcommand\vec{\mathbf}

\newcommand\bb{\begin{eqnarray}}
\newcommand\ee{\end{eqnarray}}
\newcommand{\dan}[1]{#1}

\newcommand{\changed}[1]{{\leavevmode\color{black}{#1}}}

\title{Experimental observations and modeling of sub-Hinze bubble production by turbulent bubble break-up}

\author{Daniel J. Ruth\aff{1},
Aditya K. Aiyer\aff{1},
Aliénor Rivière\aff{1,2,3},
Stéphane Perrard\aff{1,2,3},
 \and Luc Deike\corresp{\email{ldeike@princeton.edu}}\aff{1,4}}

\affiliation{\aff{1}Department of Mechanical and Aerospace Engineering, Princeton University
\aff{2}LPENS, Département de Physique, Ecole Normale Supérieure, PSL University, 75005 Paris, France
\aff{3} Physique et Mécanique des Milieux Hétérogènes, CNRS, ESPCI Paris, University PSL, Paris 75005, France
\aff{4}High Meadows Environmental Institute, Princeton University}

\begin{document}

\maketitle

\begin{abstract}
    We present experiments on large air cavities \changed{spanning a wide range of sizes relative to the} Hinze scale $\dH$, \dan{the scale} at which turbulent stresses are balanced by surface tension, disintegrating in turbulence. For cavities with initial sizes $d_0$ much larger than $\dH$ \changed{(probing up to $\dodH = 8.3$)}, the size distribution of bubbles smaller than $\dH$ follows $N(d) \propto d^{-3/2}$, with $d$ the bubble diameter. The capillary instability of ligaments involved in the deformation of the large bubbles is \changed{shown visually} to be responsible for the creation of the small ones. Turning to dynamical, three-dimensional measurements of individual break-up events, we describe the break-up child size distribution and the number of child bubbles formed as a function of $\dodH$. \changed{Then, to model the evolution of a population of bubbles produced by turbulent bubble break-up, we propose a population balance framework in which break-up involves two physical processes: an inertial deformation to the parent bubble that sets the size of large child bubbles, and a capillary instability that sets the size of small child bubbles}. A Monte Carlo approach is used to construct the child size distribution, with simulated stochastic break-ups constrained by our experimental measurements and the understanding of the role of capillarity in small bubble production. This approach reproduces the experimental time evolution of the bubble size distribution during the disintegration of large air cavities in turbulence.
\end{abstract}


\section{Introduction}

\subsection{Broader context}

Gas bubbles dispersed in liquids provide surface area through which mass can be exchanged by diffusion. Ocean-atmosphere exchanges of \ce{CO2}, for example, are enhanced  by bubble-mediated transfer in regions of the globe where high winds lead to high rates of wave breaking, as entrained air cavities break apart into small bubbles in the turbulent field under the breaking wave \citep{Deike2018,Reichl2020,Deike2022}. Further, many industrial processes involve facilitating gas transfer to a liquid through bubble interfaces \citep{Schludieter2021}. In both environmental and industrial scenarios, the breakage of bubbles by the turbulence of the bulk flow increases the total surface area through which transfers may occur and modulates the bubbles' dynamics. 

Despite the ubiquity of bubble break-up across disciplines, the physics of bubble breaking in turbulence remains to be fully understood, as turbulent effects are often accompanied by buoyant effects and shear in the mean structure of the flow \citep{Risso1998}. Further, the fast dynamics of bubble pinching have, until recently, been difficult to measure experimentally, leaving open questions regarding the final portion of the break-up process \citep{Ruth2019}. These various challenges have led to a wide variability in the predictions of models for both the rate at which bubbles break and the sizes of bubbles they break into. 

\subsection{Bubble break-up in turbulence}

We consider the break-up of a bubble with an effective diameter $d_0$, taken to be the diameter of a sphere with the same volume. Before considering the turbulent nature of the liquid around it, the bubble in a liquid is described by the density of the liquid and gas phases, $\rho$ and $\rho_\mathrm{g}$, their viscosities $\mu$ and $\mu_\mathrm{g}$, the acceleration due to gravity $g$, and the surface tension of the liquid-gas interface $\sigma$. When the carrier flow in which the bubbles are dispersed (with velocity $\vec{u}$) is turbulent, it is characterized by the dissipation rate of the turbulence $\epsilon$, which is the rate at which kinetic energy in turbulent fluctuations is dissipated to heat. The turbulence is comprised of fluctuating motions existing over a range of length scales, extending from larger motions near the integral length scale $L_\mathrm{int}$ (beyond which the velocity field becomes uncorrelated) down to the Kolmogorov scale $\eta$, at which turbulent motions are dissipated by the viscosity of the fluid \citep{Pope2000}. 

With nine independent physical parameters which span three physical dimensions, we require six dimensionless parameters to describe the problem of bubble break-up in turbulence, for which we choose
\begin{gather}
    \changed{\mathrm{We}_0 = \frac{C_2 \rho \epsilon^{2/3} d_0^{5/3}}{\sigma}}, \qquad \frac{d_0}{L_\mathrm{int}}, \qquad \frac{d_0}{l_\mathrm{cap}} = \sqrt{\frac{\rho g d_0^2}{\sigma}}, \nonumber \\
    \mathrm{Re}_\mathrm{t} = \frac{\rho L_\mathrm{int} u'}{\mu}, \qquad \frac{\rho}{\rho_\mathrm{gas}}, \qquad \frac{\mu}{\mu_\mathrm{gas}}, \label{eq:dimensionless_numbers}
\end{gather}
where the subscript "0" indicates a quantity refers to an initial condition.  The \dan{size of the parent bubble relative to the capillary length scale $l_\mathrm{cap} = \sqrt{\sigma / (\rho g)}$} describes the relative importance of gravity and surface tension effects for the parent bubble. The large-scale turbulence Reynolds number $\mathrm{Re}_\mathrm{t}$ represents the separation of length scales in the turbulence. The bubble size relative to the integral length scale $d_0/L_\mathrm{int}$, along with $\mathrm{Re}_\mathrm{t}$, describes the spatial separation between the bubble and the turbulence scales. With $\mathrm{Re}_\mathrm{t} \gg 1$ and $\rho/\rho_\mathrm{gas}$ and $\mu/\mu_\mathrm{gas}$ both fixed constants $\gg 1$ for common liquid-gas configurations, we will neglect their impact in the rest of the experimental study. \changed{The Weber number of the parent bubble $\mathrm{We}_0$, which parameterizes the balance between turbulent stresses and surface tension, will be the main parameter of focus.}

For a bubble in the inertial subrange of the turbulence ($\eta \ll d_0 \ll L_\mathrm{int}$), the ratio of the inertial stresses arising from velocity gradients in the turbulence and surface tension stresses defines the Weber number, $\mathrm{We}(d) = C_2 \rho \epsilon^{2/3} d^{5/3} / \sigma$, with $C_2=2$, and is central in the analysis of bubble break-up \citep{Risso1998,Riviere2021jfm,Perrard2021BubbleFlow}. The definition of a critical Weber number for break-up $\mathrm{We}_\mathrm{c}$ yields the Hinze scale \citep{Hinze1955},
\begin{equation}
    \dH = \left(\frac{\mathrm{We}_\mathrm{c}}{2}\right)^{3/5} \left( \frac{\sigma}{\rho} \right)^{3/5} \epsilon^{-2/5}, \label{eq:Hinze_scale}
\end{equation}
\changed{and we typically use the ratio $d/\dH = (\mathrm{We}/\mathrm{We}_\mathrm{c})^{3/5}$ in place of $\mathrm{We}$}. Estimations of $\mathrm{We}_\mathrm{c}$ vary, and generally involve either considerations of how likely a bubble is to break apart over some physically-relevant time or within some spatial observation window \citep{Hinze1955,Martinez-Bazan1999a,Risso1998,Riviere2021jfm}, or considerations of the shape of the bubble size distribution resulting from break-ups \citep{Deane2002}. Since $\mathrm{We}_\mathrm{c}$ is influenced by factors like the buoyancy and specificity of the turbulent flow, and since the turbulent stresses on a bubble are stochastic in nature, the Hinze scale as defined in \cref{eq:Hinze_scale} represents a soft limit for break-up. Different experimental and computational \dan{setups will} lead to a range of reported or inferred critical Weber numbers, which typically vary from 1---5 \citep{Riviere2021jfm,Risso1998,Hinze1955,Martinez-Bazan1999a,Vejrazka2018}. In this paper, we will use $\mathrm{We}_\mathrm{c} = 1$, consistent with our results and similar experiments in a turbulent flow forced by underwater pumps \citep{Vejrazka2018}. We note that the inertial stresses on a bubble that arise from the \dan{velocity slip} between the bubble and the surrounding liquid can induce stresses comparable to those \dan{associated with the turbulence's inherent} velocity gradients at the bubble scale \citep{Masuk2021}, that eddies smaller than the bubble can also contribute to deformation and break-up \citep{Luo1996,Qi2022}, and that the turbulent flow can trigger bubble shape oscillations \citep{Risso1998,Ravelet2011}. These factors will contribute to bubble deformation and break-up in ways that are not directly parameterized in the definition of $\dH$.


\changed{The bubble size distribution $N(d)$ gives the number density of bubbles with diameter $d$, and given the nature of experiments reported in this paper, we define it such that $N(d) \mathrm{d} d$ gives the total number of bubbles with diameters $\in (d,d+\mathrm{d}d)$.} \cite{Garrett2000} proposed that, for bubbles larger than the Hinze scale, a power-law scaling $N(d) \propto d^{-10/3}$ describes the steady-state bubble size distribution, assuming that the break-up rate scales with the turbulent frequency at the bubble size. This regime has since been reported in several experiments \citep{Deane2002,Blenkinsopp2010,Rojas2007} and simulations \citep{Deike2016,Wang2016,Gao2021BubbleCrests,Chan2021,Soligo2019,Riviere2021jfm,Mostert2021}. For smaller bubbles, the size distribution typically exhibits a shallower slope \citep{Deane2002,Blenkinsopp2010}, with fewer studies resolving this range of scales and some variation in the values that have been reported. The $N(d) \propto d^{-3/2}$ distribution for $d<\dH$ has been observed experimentally \citep{Deane2002} and numerically \citep{Wang2016,Mostert2021} for bubbles under breaking waves, though the identification of a sub-Hinze power-law slope is additionally complicated by the transient nature of bubble disintegration \citep{Riviere2021jfm} and breaking wave \citep{Mostert2021} events. Recent work has identified the capillary pinching of gas ligaments created by turbulent deformations as an origin of sub-Hinze bubbles, with theoretical arguments relating to the timescale over which such pinching occurs supporting the $N(d) \propto d^{-3/2}$ sub-Hinze scaling \citep{Riviere2021cap}. Relating measured size distributions to theoretical scalings derived from break-up physics is complicated by the fact that bubbles' motions, and hence their residence time in some experimental domain, are dependent on their size and the characteristics of the turbulence they encounter \citep{Garrett2000}. \dan{Smaller bubbles or bubbles in regions of more intense turbulence will rise slower than others}, for example \citep{Ruth2021}; accounting for these effects requires detailed knowledge of the size dependencies of the bubbles' motions.

\subsection{Child size distribution and break-up time scales}

In this work, we will employ experimental observations to describe bubble break-up over a range of spatial scales: we consider parent bubbles ranging in size from the Hinze scale to \changed{$d_0 = 8.3 \dH$}, and investigate how they break up to produce child bubbles that may be orders of magnitude smaller than the Hinze scale. As volume is conserved in any break-up, we will work with bubble volumes $V= \pi d^3 / 6$ when discussing bubble break-up, denoting parent bubble volumes by $V=\Delta$ and child bubble volumes by $V=\delta$. 

Expressions for a break-up kernel $f(\delta;\Delta)$, for which $f(\delta;\Delta) \mathrm{d} \delta$ gives the rate at which a parent bubble of volume $\Delta$ will break into a child bubble with volume $\in (\delta,\delta+\mathrm{d}\delta)$ in some turbulent flow, are informed by experiments and simulations on break-up. Most experimental studies have involved air bubbles in water under Earth's gravitational acceleration, with turbulence in the water generated by one or more jets \citep{Martinez-Bazan1999a,Vejrazka2018,Qi2020}, rotating blades \citep{Ravelet2011}, or by turbulent flow through a reactor or channel \citep{Andersson2006}. \cite{Risso1998} performed experiments on bubble break-up in microgravity to remove the effects of buoyancy, which also contributes to bubble deformation and break-up, and more recently, \cite{Riviere2021jfm} performed direct numerical simulations (DNSs) of bubble break-up without gravity, solving the full two-phase Navier Stokes equations for a bubble subjected to homogeneous, \dan{isotropic} turbulence. 

These studies have confirmed that the time over which a break-up occurs is controlled by both the turbulent scales and the bubble's oscillatory scales. \cite{Riviere2021jfm} showed that, as a bubble of size $d_0 \gg \dH$ is introduced to turbulence, it first breaks up after a time comparable to eddy turn-over time at its scale, $T_\mathrm{turb}(d_0) = \epsilon^{-1/3} d_0^{2/3}$. Experimental studies have shown that the time over which deformation occurs prior to break-up scales similarly \citep{Qi2020,Risso1998}. As the deformation of moderately-sized bubbles is also impacted by the surface tension, capillary dynamics remain important, as a bubble's natural oscillation \dan{frequency remains} apparent in its shape oscillations \citep{Risso1998,Ravelet2011,Perrard2021BubbleFlow}. Further, the turbulent turnover time is typically comparable to the capillary oscillation time at the parent bubble scale for air bubbles in water at moderate $\dodH$, which can lead to a resonance which aides break-up \citep{Risso1998,Ravelet2011}. 

The break-up frequency $\omega$ is defined as the inverse of the typical time until a bubble undergoes a break-up, \changed{and is distinct from the (necessarily shorter) typical duration over which a break-up occurs}. \cite{Ravelet2011} showed that the distribution of the times until a bubble breaks mirrors the distributions of the times between severe shape deformations and the times between large instantaneous Weber numbers. The most energetic scales capable of deforming a bubble are those at the scale of the bubble, and experiments from which $\omega$ was extracted suggested that the break-up frequency initially increases with bubble size as the turbulence becomes more capable of counteracting surface tension, and then decreases for even larger bubbles, as the time required for a turbulent eddy to act across the bubble scale becomes longer \citep{Martinez-Bazan1999a}, though this analysis may have missed break-ups in which one child bubble size is close to the parent size \citep{Lehr2002}. Recent experiments from \cite{Qi2022} showed that eddies smaller than $d_0$ can also cause break-up, and other theoretical analyses have considered the action of a range of turbulent scales which may cause break-up. In such models, the product of the rate at which eddies of a given size interact with a bubble and each interaction's likelihood of causing break-up are integrated over a range of eddy sizes \citep{Prince1990,Tsouris1994,Luo1996,Lehr2002,Aiyer2019,Yuan2021}, causing the break-up frequency to increase with the bubble size as more turbulent scales contribute to break-up.

Various models for the child size distributions $p(\delta;\Delta)$ have been proposed, most of which assume that each break-up produces two bubbles. The child size distribution has been described with a $\cap$--shaped dependence on $\delta$---that is, the most likely outcome is to produce child bubbles that are comparable in \dan{size} to the parent bubble \citep{Martinez-Bazan1999b,Martinez-Bazan2010}; or with a $\cup$-- or W--shaped child size distributions, in which small bubbles are more likely to be produced than moderately-sized ones \citep{Qi2020,Riviere2021jfm,Vejrazka2018,Andersson2006,Tsouris1994,Luo1996,Lehr2002,Yuan2021,Qi2020}. Experimental and numerical evidence suggests that break-ups often produce just two child bubbles when $\dodH$ is close to 1 \citep{Vejrazka2018,Riviere2021jfm}. However, break-ups at larger $\dodH$ are more severe and often result in more than two child bubbles being formed in a single coherent event \citep{Vejrazka2018,Hinze1955,Riviere2021jfm}. \cite{Hill1996} developed generalized expressions for $p(\delta;\Delta)$ as products of power-law relations (each $\propto \delta^\alpha$) for $\alpha>-1$ and integer numbers of child bubbles, which by design satisfy constraints relating to the sizes of the bubbles formed. Their analysis was extended to \dan{break-ups with a non-integer average number of child bubbles} by \cite{Diemer2002}.

In the work discussed so far, the role of capillarity has been to counteract the turbulent stresses and prevent severe deformation, while also providing a resonance mechanism at moderate $\dodH$. However, more recent work has shown that capillarity also plays an important role late in the break-up process, even after a turbulent stress has decidedly overcome it. \cite{Andersson2006} showed that asymmetries in a deformed bubble shape can become more pronounced as a bubble breaks apart due to the variation in capillary pressure associated with the deformation. More recently, \cite{Riviere2021cap} showed that very small bubbles originate not from turbulent motions at very small scales, but rather from the capillary instabilities of ligaments arising from much larger-scale deformations. 

\subsection{Outline of the paper}

\label{sec:problem_characterization}


\changed{In this work we address the problem of bubbles breaking up in forced turbulence, which is applicable to break-up under breaking waves and in industrial reactors. We probe a wide range of scales, with bubbles ranging in size from the Hinze scale to $d=8.3 \dH$ (corresponding to $\mathrm{We}_0 = 34$). Further, we resolve the size distribution down to approximately an order of magnitude smaller than $\dH$, enabling us to identify the way in which the sub-Hinze size distribution scales when there is a large separation between the Hinze scale and the bubbles which break.} 

The experiment set-up, including the turbulence generation, is detailed in \Cref{sec:experiment}. The results on the disintegration of large air cavities are given in \Cref{sec:air_cavity_results}, spanning a wide range of $d_0/d_H$. We demonstrate experimentally that a $N(d)\propto d^{-3/2}$ distribution below the Hinze scale is observed when the initial cavity size is much larger than the Hinze scale, supporting the notion that the capillary pinching dynamics proposed by \cite{Riviere2021cap} are effective at producing sub-Hinze bubbles. The dynamically-tracked individual bubble break-ups with moderate $\dodH$ and resulting child size distributions are discussed in \Cref{sec:dynamical}. In \Cref{sec:model} we develop a model for turbulent bubble break-up that unifies the turbulent inertial dynamics with the faster, capillary pinching dynamics responsible for sub-Hinze bubble production, \changed{ascribing these physical mechanisms to various components of a modeled child size distribution}. The model is informed by both experimental observations of the disintegrations of air cavities of various sizes and by experimental \dan{and numerical} observations of individual break-up events. Concluding remarks are given in \Cref{sec:breakup_conclusions}.


\section{Experimental setup}
\label{sec:experiment}

This paper presents the results of two separate, complementary experiments, both involving air bubbles breaking apart in forced water turbulence. In the first, we generate large cavities of air with sizes much larger than the Hinze scale (with $\dodH$ betwen 2.1 and 8.3) and measure the transient evolution of the bubble size distribution as the cavity disintegrates in successive break-ups. In the second experiment, we introduce moderately-sized bubbles (with $\dodH$ between 1 and 3) into the turbulence, and track the outcomes of their individual break-ups. The turbulence generation is identical in both set-ups.

\subsection{Turbulence generation and characterization}

\begin{figure}
    \centering
    \begin{minipage}{.4\textwidth}
        \centering
        \begin{overpic}[width=1\linewidth]{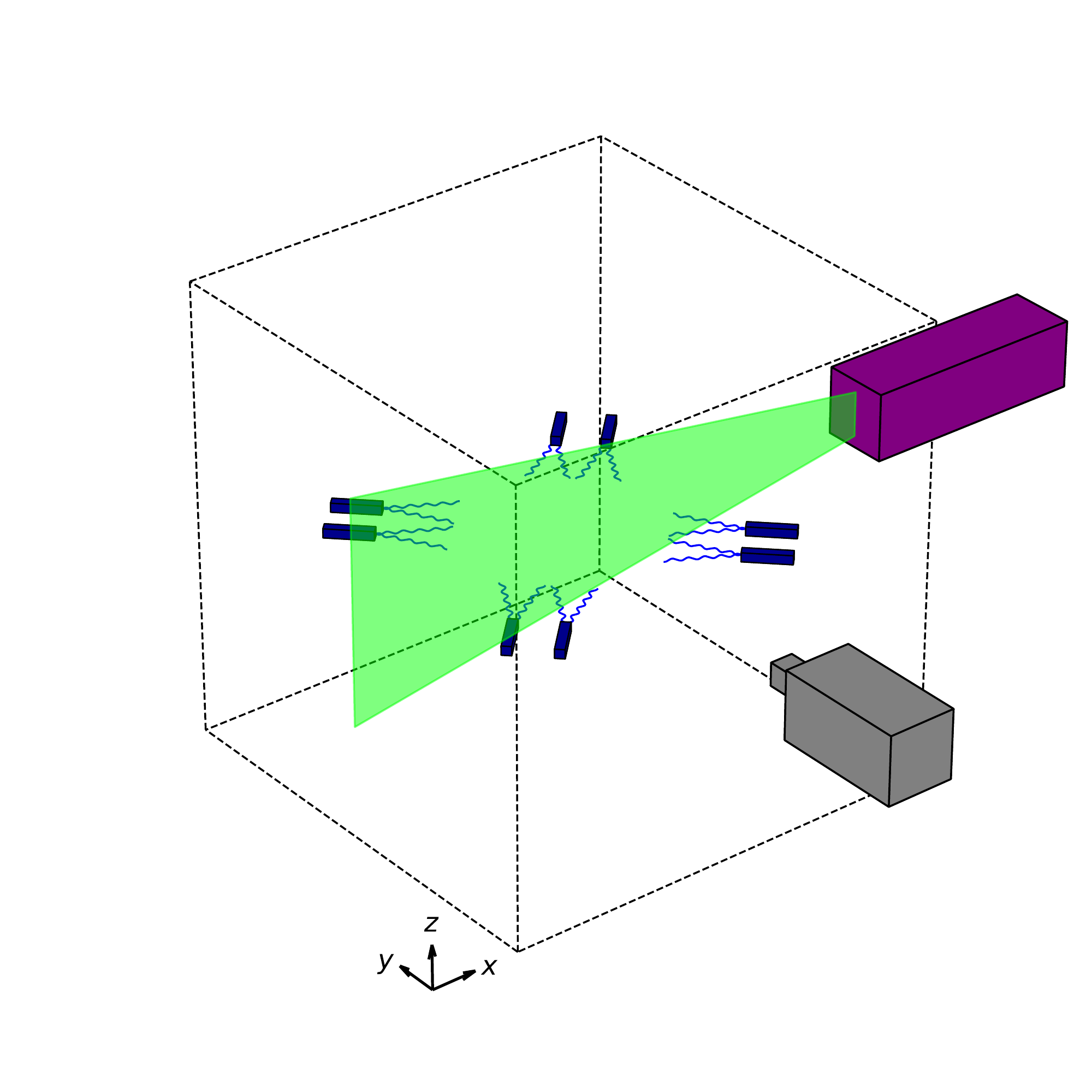}
        \put(0,85){(a)}
        \end{overpic}
    \end{minipage}%
    \begin{minipage}{0.5\textwidth}
        \centering
        \begin{overpic}[width=0.95\linewidth]{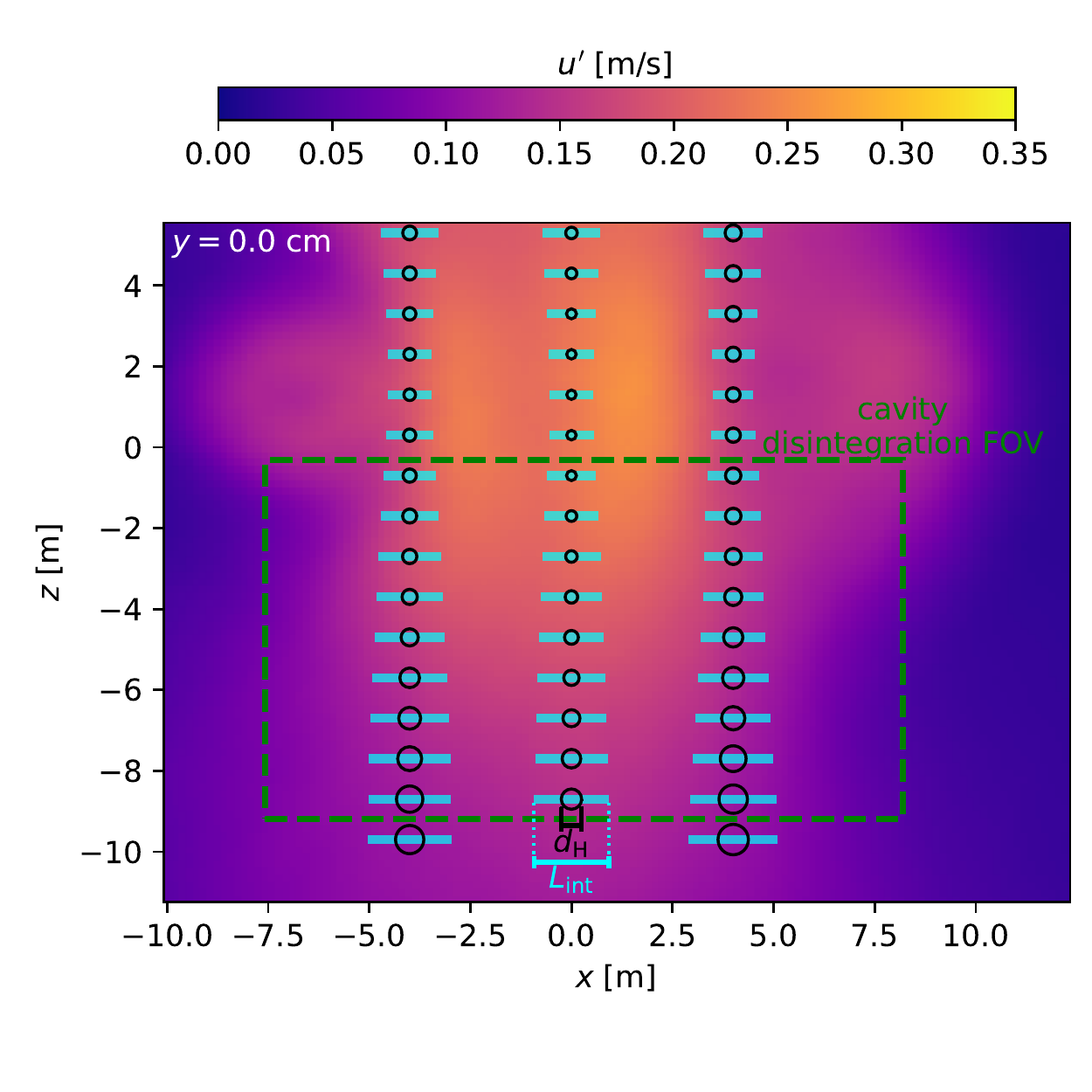}
        \put(0,79.5){(b)}
        \end{overpic}
    \end{minipage}
    \caption{The turbulence generation and characterization. (a) A sketch of the experiment (not to scale), consisting of a \SI{0.37}{m^3} tank of water in which four pumps, each split to two outlets, are arranged at the corners of a square in the horizontal plane. The turbulence is characterised with particle image velocimetry performed separately in nine parallel planes, with illumination provided by a laser sheet (shown in green). (b) Properties of the turbulent flow field in the central plane of the experiment. The background shows the local value of $u'$, denoted by the color given in the colorbar. The green dashed rectangle shows the field of view employed in the large air cavity disintegration experiments. The diameter of the black circles denotes the Hinze scale $d_\mathrm{H}$ at various $x$ and $z$. The length of the cyan rectangles denotes the integral length scale $L_\mathrm{int}$ at those locations. }
    \label{fig:exp_and_PIV_singleplane}
\end{figure}

Turbulence in a \SI{0.37}{m^3} water tank is generated by the convergence of eight turbulent jets created by four submerged water pumps, as sketched in \Cref{fig:exp_and_PIV_singleplane} (a) and described in greater detail in \cite{Ruth2021}. The flow from each pump is split into two parallel jets at a Y, with each outlet separated by \SI{7.8}{cm}, with the centers of the Y forming the vertices of a \SI{25}{cm} square in the horizontal plane. \Cref{fig:exp_and_PIV_singleplane} (b) presents properties of the flow as characterized in the central plane ($y=0$) of the experiment with two-dimensional, two-component particle image velocimetry (PIV). The background gives the local fluctuation velocity $u' = \sqrt{({u'_x}^2+{u'_z}^2)/2}$, where $u'_i=\sqrt{\overline{(u_i-\overline{u_i})^2}}$ and overbars denote averaging in time. $u'$ tends to be largest in the plane of the jets ($ z \approx \SI{0.01}{cm}$) and in the region below their convergence zone ($x \approx y \approx 0$). PIV is performed in nine parallel planes, enabling the three-dimensional interpolation of turbulence quantities at any location within the measurement domain.

As described in \cite{Ruth2021}, we compute the integral length scale $L_\mathrm{int}$ locally at each point in the flow by integrating the spatial autocorrelation function. It changes throughout the experiment, being the shortest where the turbulence is the strongest. The cyan lines in \Cref{fig:exp_and_PIV_singleplane} (b) denote the value of $L_\mathrm{int}$ at various locations in the central plane of the experiment: $L_\mathrm{int}$ is shortest near the convergence of the jets, and grows at lower and higher depths. With $u'$ and $L_\mathrm{int}$ calculated from the PIV data, we can then compute the local turbulence dissipation rate \dan{under the assumption of isotropy} with $\epsilon = C_\epsilon u'^3/L_\mathrm{int}$, with $C_\epsilon = 0.7$ \citep{Sreenivasan1997}, and the Kolmogorov microscale with $\eta = ((\mu/\rho)^3/\epsilon)^{1/4}$ \citep{Pope2000}. The Hinze scale $\dH$, calculated using \cref{eq:Hinze_scale}, is denoted at various locations by the diameter of the black circles drawn in \Cref{fig:exp_and_PIV_singleplane} (b). The Hinze scale is smaller where the turbulence is more intense, meaning that more bubbles will be larger than the Hinze scale and susceptible to break-up at these locations. We refer to \cite{Ruth2021} for more details on the structure of the turbulence field and for maps of turbulent quantities outside of the central plane.

\subsection{Large cavity disintegration experiment}
\label{sec:exp_cavity}

\begin{figure}
    \centering
    \begin{minipage}{.4\textwidth}
        \centering
        \begin{overpic}[width=1\linewidth]{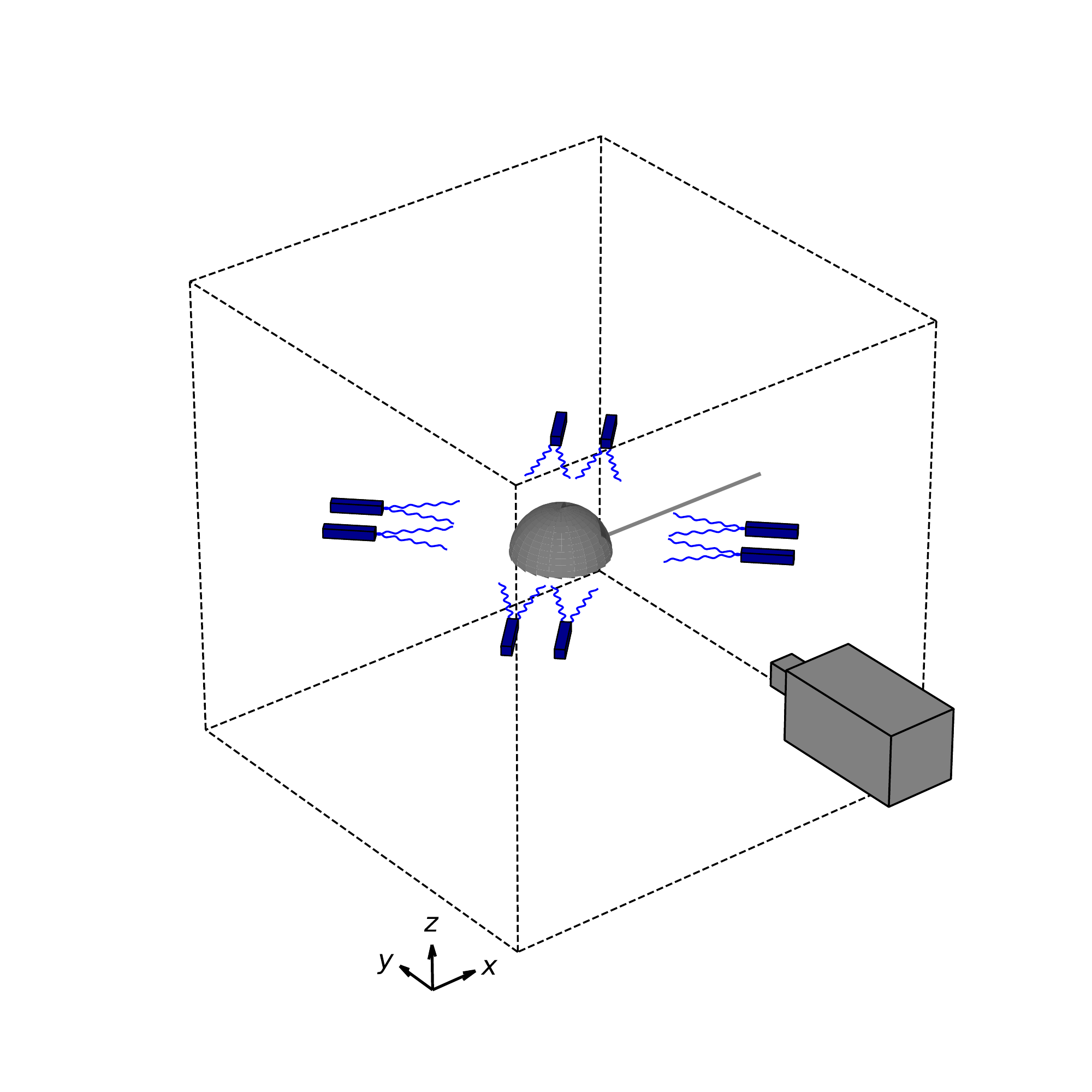}
        \put(0,86){(a)}
        \end{overpic}
    \end{minipage}%
    \begin{minipage}{0.5\textwidth}
        \centering
        \begin{overpic}[width=1\linewidth]{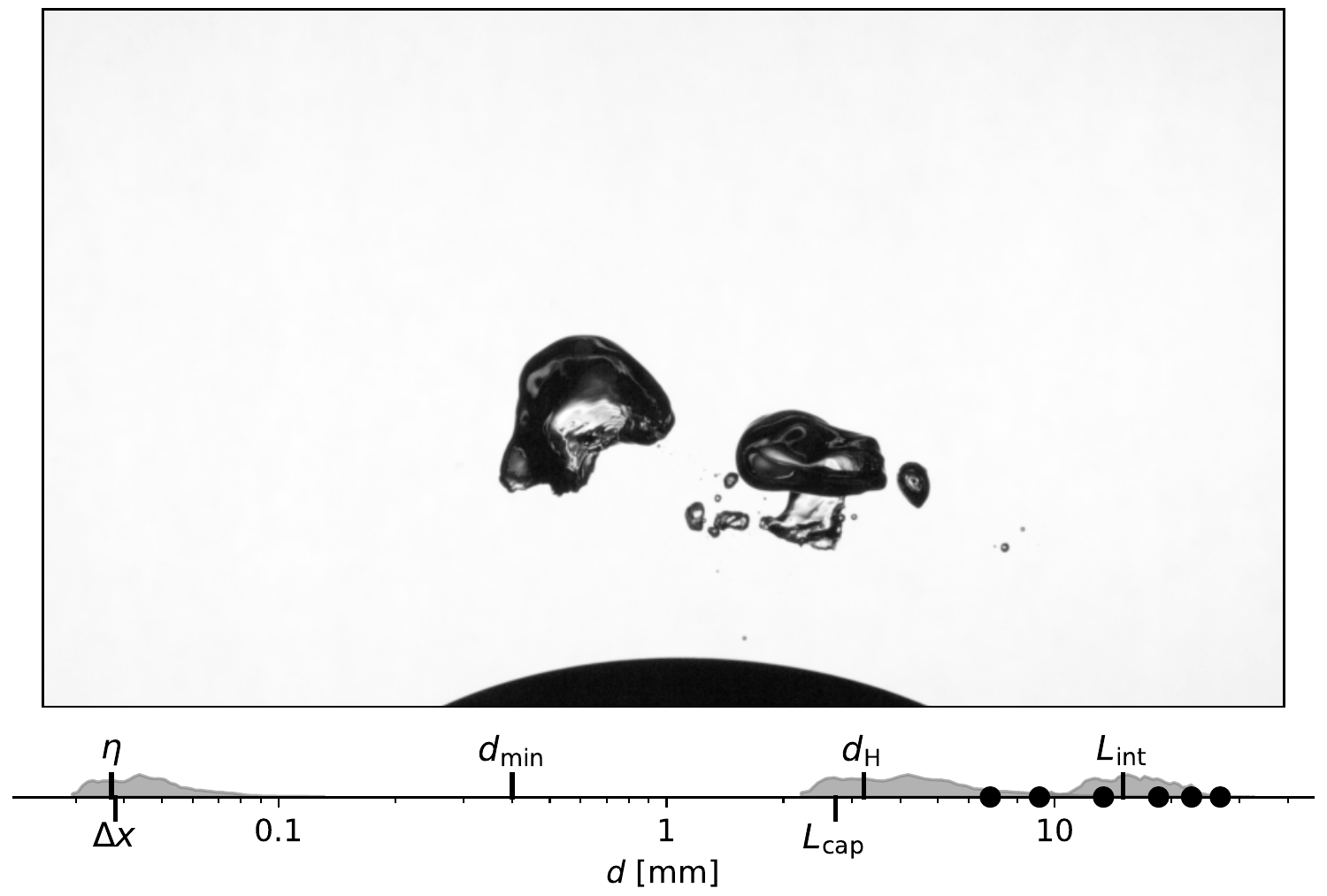}
        \put(4,62){(b)}
        \put(-6,5.5){(c)}
        \end{overpic}
    \end{minipage}
    \caption{The experiment on large cavity disintegration. (a) Schematic of the experiment. Air is bubbled into an inverted hemispherical cup located just under the convergence of the turbulent jets, and the cup is rapidly rotated to expose the air to the turbulence. The experiment is lit from behind (not shown) and filmed with a high-speed camera. (b) One representative image of a cavity breaking apart, with the cup still slightly visible at the bottom of the image. (c) The characteristic length scales $\eta$, $\dH$, $L_\mathrm{cap}$, and $L_\mathrm{int}$ taken in analyzing the data, the pixel size $\oldDelta x$ and the minimum bubble size considered $d_\mathrm{min}$, and the diameters of the air cavities studied (circles). Distributions of the turbulence quantities in the field of view in the center of the tank (within the green rectangle in \Cref{fig:exp_and_PIV_singleplane}) are also given in gray.}
    \label{fig:schematic_cavity}
\end{figure}

For the experiment on large cavity break-ups, air cavities were produced following \cite{Landel2008} by placing a hollow hemispherical cup with $R=\SI{5}{cm}$ underwater, sketched in \Cref{fig:schematic_cavity} (a), and bubbling a known volume of air $V_0 = \pi d_0^3 / 6$ into it. Once bubbles in this cup have coalesced into a single air cavity, the cup is then inverted by rotating it rapidly half a revolution, such that the air inside is suddenly no longer constrained by the curved cup surface. The top surface of the initial volume of air roughly conforms to the curved inner surface of the cup. The large air cavity, having been suddenly exposed to stresses from the surrounding turbulence and its buoyant rise through the water, deforms and starts a complex sequence of break-ups, leading to its disintegration. The surface of the cup rotates with a speed around \SIrange{0.4}{0.9}{m/s}, and we have checked that this speed does not systematically impact the early stages of the bubble size distribution. \changed{Further, similar experiments run without turbulence yield very little break-up, as evidenced in \Cref{sec:quiescent_breakup_comparison}.}

The turbulent flow in the region of the tank imaged in this experiment is denoted by the green rectangle in \Cref{fig:exp_and_PIV_singleplane} (b). The turbulence varies spatially, so to simplify the analysis, we take $u' \approx \SI{0.2}{m/s}$, $L_\mathrm{int} \approx \SI{1.5}{cm}$, and $\eta \approx \SI{37}{\micro m}$ as characteristic values, which set $d_\mathrm{H} = \SI{3.2}{mm}$ and $\mathrm{Re}_\mathrm{t} = 3400$.  These length scales are denoted in \Cref{fig:schematic_cavity} (c), which also gives the distribution of the length scales present in the field of view in the middle of the tank. The mean flow is downwards with $\overline{W} \approx \SI{-0.25}{m/s}$, largely counteracting the buoyant rise speed of larger bubbles. This enables the bubble population to linger in the measurement region for a sufficient period of time to image it over multiple large-scale eddy turn-over times $T_\mathrm{int} = L_\mathrm{int}/u' \approx \SI{0.075}{s}$. The cavities range in size between $\dodH = 2.12$ and 8.30. Data for each condition, as well as the number of runs recorded at each, are given in \Cref{tab:cavity_conditions}. 

\begin{table}
  \begin{center}
\def~{\hphantom{0}}
\begin{tabular}{l|ccccccc}
experiment & $d_0$ [\si{cm}] & runs & $\mathrm{We}_0$ & $\dodH$ & $d_0/\eta$ & $d_0 / L_\mathrm{int}$ & $d_0/l_\mathrm{cap}$ \\
\hline
cavity disintegration& 0.68 & 20 & 3.5 & 2.12 & 184 & 0.46 & 2.51\\
& 0.91 & 15 & 5.7 & 2.84 & 247 & 0.61 & 3.36\\
& 1.34 & 15 & 10.7 & 4.15 & 361 & 0.89 & 4.91\\
& 1.85 & 15 & 18.5 & 5.76 & 500 & 1.24 & 6.80\\
& 2.25 & 10 & 25.6 & 7.00 & 608 & 1.50 & 8.28\\
& 2.67 & 11 & 34.0 & 8.30 & 721 & 1.78 & 9.81\\
\hline
individual break-ups & 0.54 $\pm$ 0.17 & 162 & 3.1 $\pm$ 1.7 & 1.89 $\pm$ 0.64 & 156 $\pm$ 50 & 0.41 $\pm$ 0.14 & 1.99 $\pm$ 0.62\\
\end{tabular}
\caption{Conditions of the experiments. Characteristic values are given for each of the cavity disintegration cases. For the experiments on individual bubble break-up, the mean and standard deviation among the 162 recorded cases are given for each quantity.}
\label{tab:cavity_conditions}
\end{center}
\end{table}

One image is shown in \Cref{fig:schematic_cavity} (b). The cup is visible in the bottom of the image as it has not yet fully rotated out of the field of view. The measurement region, \changed{which spans \SI{15.8}{cm} in the $x$ direction and \SI{8.9}{cm} in the $z$ direction,} is illuminated from the back, and the disintegration of the cavity is filmed with a high-speed camera at \SI{500}{Hz} with a spatial resolution of \SI{38}{\micro m/pixel}. \changed{The field of view is much larger than all the bubbles considered, so it does not introduce a significant bias related to bubbles whose images extend partially outside the field of view.} Bubbles are detected with an image processing method described in \Cref{sec:cavity_bubble_detection}, and their \changed{effective diameters $d$ are determined as the equivalent diameter of a circle with the same area as the projected bubble image}. In analyzing the data, we consider only bubbles for which $d\geq \SI{400}{\micro m}$, for which the detection is less sensitive to the chosen image intensity threshold. Given the typical severe deformation and overlapping images of larger bubbles ($d \gtrsim \SI{6}{mm}$), we note that their sizes will tend to be over-estimated by this method. The air void fraction in the vicinity of the cavity is high enough that we are unable to track the dynamics of individual break-ups, so we restrict our analysis to the resulting bubble size distribution.

To account for the limited field of view in our experiments, we adjust the measured size distributions by keeping a record of bubbles which have left and entered the field of view. Those which leave are "locked" into the bubble record used in computing the size distributions, while those that enter the field of view are excluded from the calculation of the size distribution. This, along with a slight smoothing in $d$ and $t$ to account for the limited number of bubbles observed at early times or with small cavities, is described in greater detail in \Cref{sec:advection_correction}, and has only a limited impact on the results reported in this paper, as we do not consider the size distribution at late times.

\subsection{Individual break-up tracking experiment}
\label{sec:exp_dynamical}

\begin{figure}
    \begin{minipage}{.4\textwidth}
        \centering
        \begin{overpic}[width=1\linewidth]{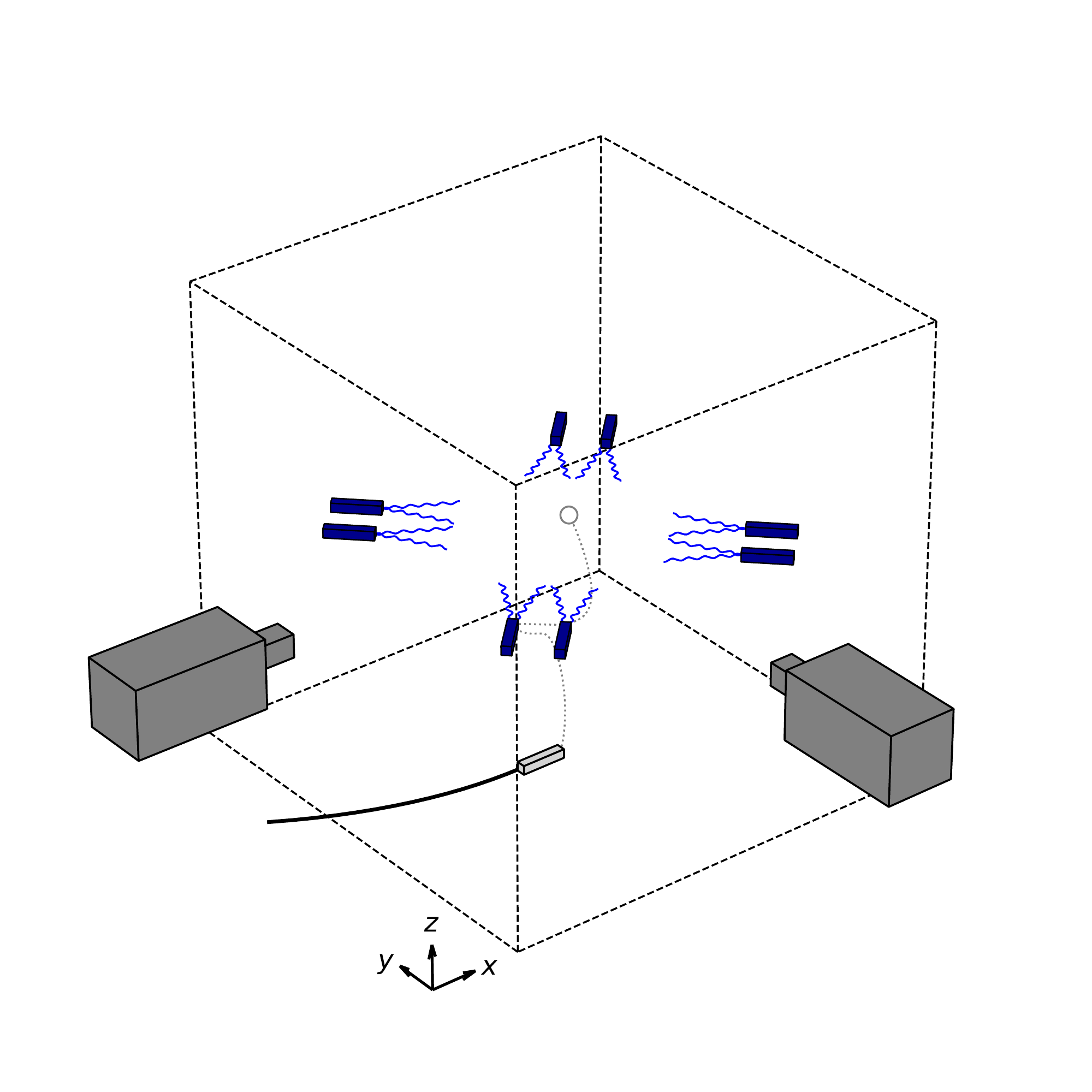}
        \put(0,85){(a)}
        \end{overpic}
    \end{minipage}%
    \begin{minipage}{0.5\textwidth}
        \centering
        \begin{overpic}[width=0.95\linewidth]{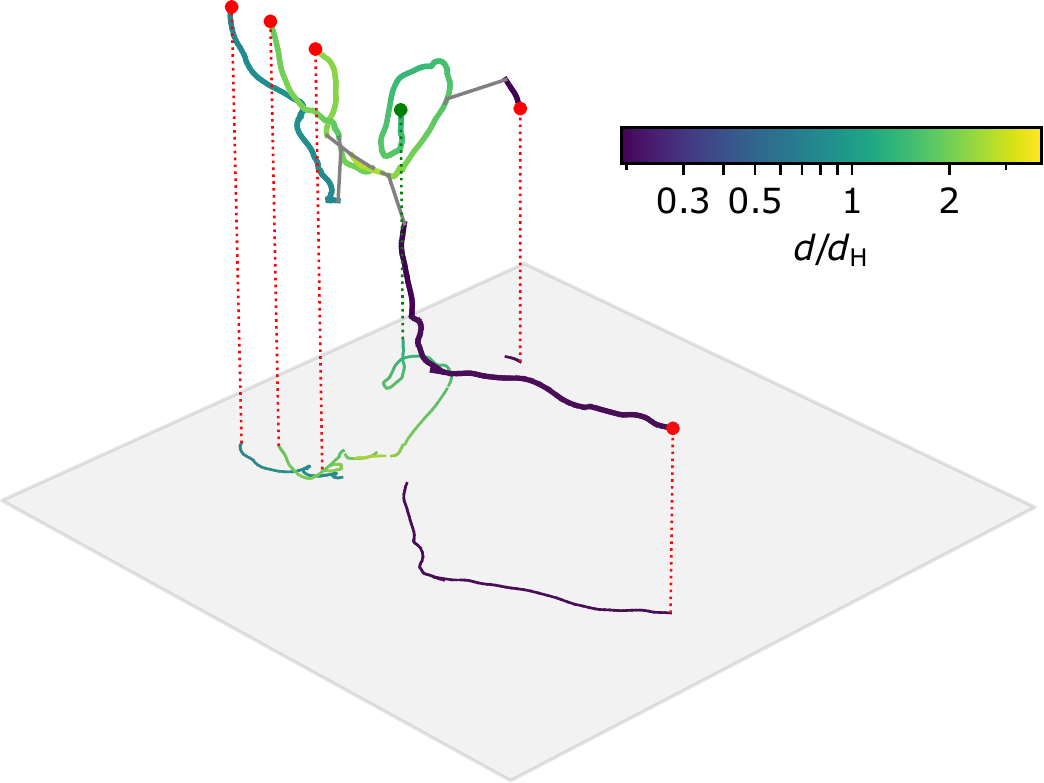}
        \put(0,66){(b)}
        \end{overpic}
    \end{minipage}
    \caption{The experiment to obtain dynamic reconstructions of individual break-up events. (a) Schematic of the experiment. Air bubbles are introduced through a needle at the bottom of the tank and rise into the turbulence created by the jets. The bubbles are filmed with two high-speed cameras, enabling the determination of the 3-D bubble trajectories. (b) The trajectories of parent and child bubbles involved in one break-up event, \dan{and their projections onto the horizontal ($x-y$) plane}. The color corresponds to the bubble's size relative to the local Hinze scale, \changed{which varies spatially with $\epsilon$ as the bubble size is fixed}. \changed{The green dot denotes the first detected position of the parent bubbles; the red dots denote the final detected position of the child bubbles.}}
    \label{fig:schematic_dynamical}
\end{figure}

In the second set of experiments of bubble break-up, we dynamically track the individual break-ups of bubbles in the turbulent region. As sketched in \Cref{fig:schematic_dynamical} (a), bubbles are introduced to the bottom of the tank through a needle and rise to the turbulent region. Two cameras\changed{,} which are synchronized with a function generator\changed{,} film at \SI{1000}{fps}. They are oriented \SI{90}{\degree} from each other and their fields of view overlap in a measurement volume of approximately \SI{200}{cm^3}. The cameras are calibrated by mapping their pixels to the paths of the light rays reaching the pixels, following the method presented by \cite{Machicoane2019AImaging}. Then, following a method similar to that used in \cite{Ruth2021}, we identify the 3-D location of the bubbles which are simultaneously captured by each camera. The spatial resolution of each camera varies with the position of the bubble, but the typical value of the two cameras are \SI{28.9}{\micro m/pixel} and \SI{57.1}{\micro m / pixel}. An approximate lower bound for the size of the smallest resolved bubble is then $d_\mathrm{min} \approx \SI{200}{\micro m}$. 

The trajectories taken by the bubbles are then determined using the Python package Trackpy \citep{trackpy}, which implements the algorithm from \cite{Crocker1996}. Such trajectories are shown in \Cref{fig:schematic_dynamical} (b). Using the three-dimensional map of the turbulence statistics obtained from PIV, we compute the bubble's size relative to the local Hinze scale $\ddH$ (computed with the local value of $\epsilon$) at each bubble location, which is encoded in the color in the figure. The mean dissipation rate at the break-up locations is $\epsilon = \SI{0.52}{m^2/s^3}$, with a standard deviation of \SI{0.21}{m^2/s^3}. The mean values and standard deviations of quantities describing the initial conditions for the break-ups studied in this experiment are given in \Cref{tab:cavity_conditions}.

From the bubble trajectories, we identify each time a bubble breaks apart, which occurs when a new trajectory (or trajectories) appears in the vicinity of a previously-existing bubble. As the tracking algorithm will initially link the parent bubble to only one of the child bubbles, the parent bubble trajectory is then split at this time, and both child bubbles are treated equally. These events are denoted by the gray lines connecting the "end" of one bubble to the "beginning" of another in \Cref{fig:schematic_dynamical} (b). Given the complex deformations involved in some break-ups, the method does not always resolve the fast splitting dynamics accurately; the break-up child size distributions we report, however, are not sensitive to the order of events occurring within one break-up event.

\section{Size distribution evolution during the disintegration of a large air cavity}
\label{sec:air_cavity_results}

Here, we present experimental results on the disintegration of air cavities of various sizes from the experiment described in \Cref{sec:exp_cavity}. First, we qualitatively discuss the break-up of cavities in two illustrative cases, one close to the critical size for break-up, and one with a large separation of scales between the cavity and the Hinze scale. Then, we analyze the transient evolution of the bubble size distributions.

\subsection{Disintegration of cavities of increasing sizes}

\begin{figure}
\centering
  \includegraphics[width=0.75\linewidth]{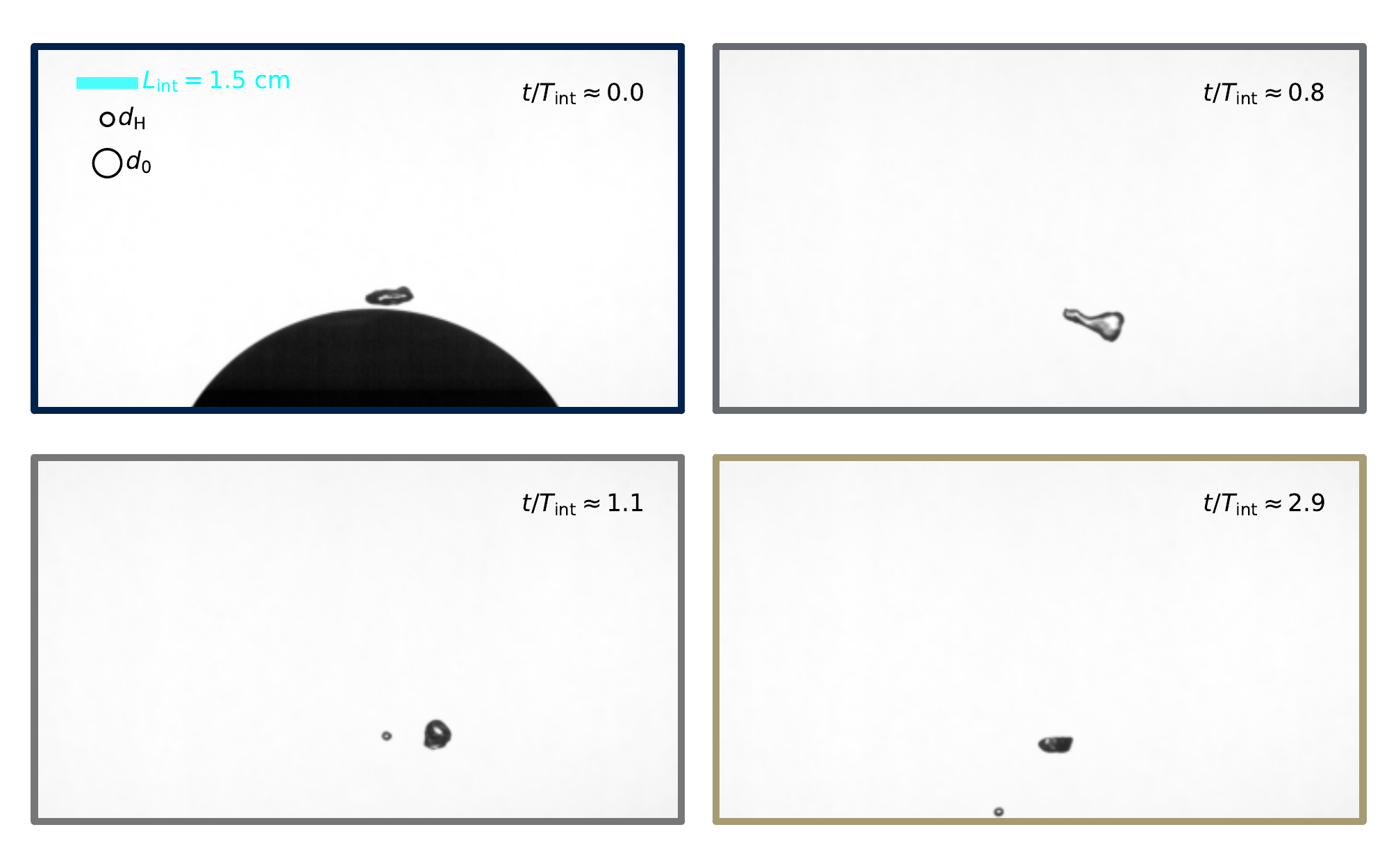}
\caption{The disintegration of an air cavity with $d_0/d_\mathrm{H} \approx 2.1$, involving just one break-up during the interval shown.}
\label{fig:bubble_release_small}
\end{figure}

\begin{figure}
\centering
  \includegraphics[width=0.75\linewidth]{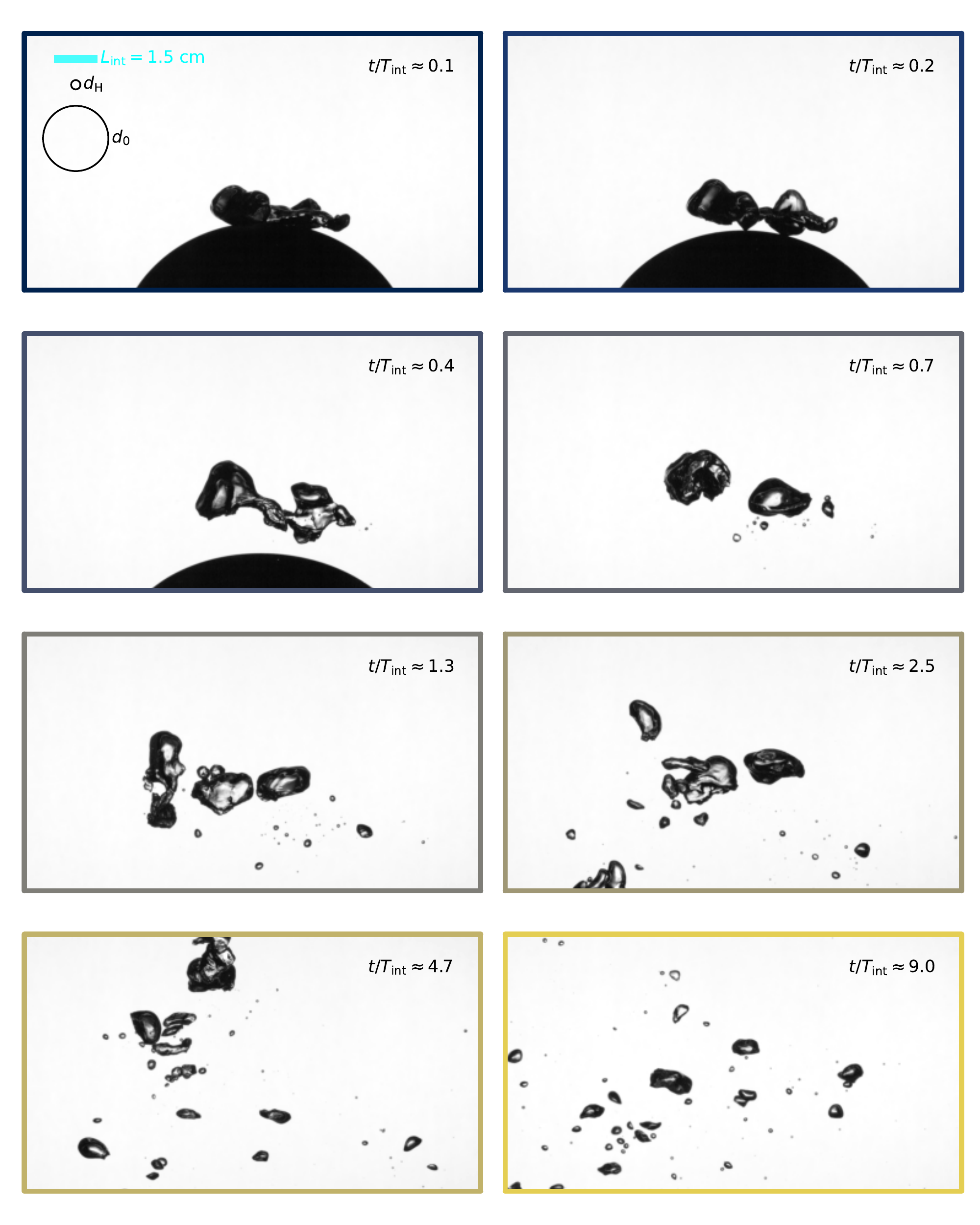}
\caption{The disintegration of an air cavity with $d_0/d_\mathrm{H} \approx 7.0$.}
\label{fig:bubble_release_big}
\end{figure}

The break-ups of two air cavities, one with $d_0 = \SI{0.68}{cm}$ and one with $d_0 = \SI{2.3}{cm}$, are shown in \Cref{fig:bubble_release_small} and \Cref{fig:bubble_release_big}, respectively. These correspond to non-dimensional sizes of $d_0/\dH  = 2.1$ and $7.0$, $d_0/L_\mathrm{int} = 0.46$ and $1.5$, and \dan{$d_0/l_\mathrm{cap} = 2.51$ and $8.28$}. As a reference, the constant values taken for $L_\mathrm{int}$ and $d_\mathrm{H}$ and the initial size of the cavity $d_0$ are denoted in the top-left corner of the first image. In both cases, the hemispherical cup which had constrained the bubble is visible at early times as it is rotated away.

In the disintegration of the smaller cavity, with $d_0/\dH = 2.1$ (shown in \Cref{fig:bubble_release_small}), the bubble emerges from the cup with a moderate deformation caused by buoyancy and the surrounding turbulence. Eventually, within approximately one integral-scale turn-over time, the bubble becomes more elongated and breaks into two bubbles. One is near the parent bubble in size, and other is slightly smaller than the Hinze scale. These two bubbles persist without breaking for at least $\sim 2$ more integral-scale turn-over times, at which point the smaller of the two bubbles is advected out of the field of view by the downwards mean flow.

The deformation to the larger cavity shown in \Cref{fig:bubble_release_big} is more severe, leading to a more complex sequence of events during its disintegration. Upon emerging from the rotating cup, the cavity is flattened due to buoyancy (as \dan{$d_0 / l_\mathrm{cap} = 8.28$} for this case), and turbulent deformations to the cavity shape on the order of the cavity size itself quickly develop. By $t/T_\mathrm{int} \approx 0.4$, the cavity consists of two lobes (each of which is significantly deformed), separated by a shrinking neck of air. By the time the neck has pinched apart ($t/T_\mathrm{int} \approx 0.7$), the two larger child bubbles stemming from the two lobes are accompanied by much smaller child bubbles (some with $d \ll d_\mathrm{H}$ and $d \ll d_0$) which were formed during the collapse of the air neck. The larger child bubbles themselves go on to further break apart in a chain of break-ups, some of which similarly involve small bubble production via the collapse of elongated air necks. Many small bubbles which are more than an order of magnitude smaller than the initial one are eventually visible. At much later times, the largest bubbles have risen out of the field of view, and the total air volume imaged is significantly decreased.

\subsection{Transient evolution of the bubble size distributions}

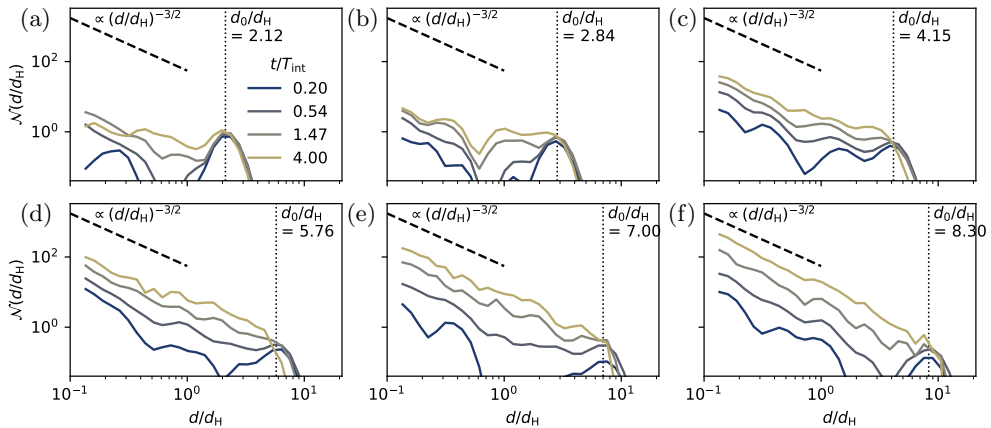
\begin{figure}
\centering
  \begin{overpic}[width=1\linewidth]{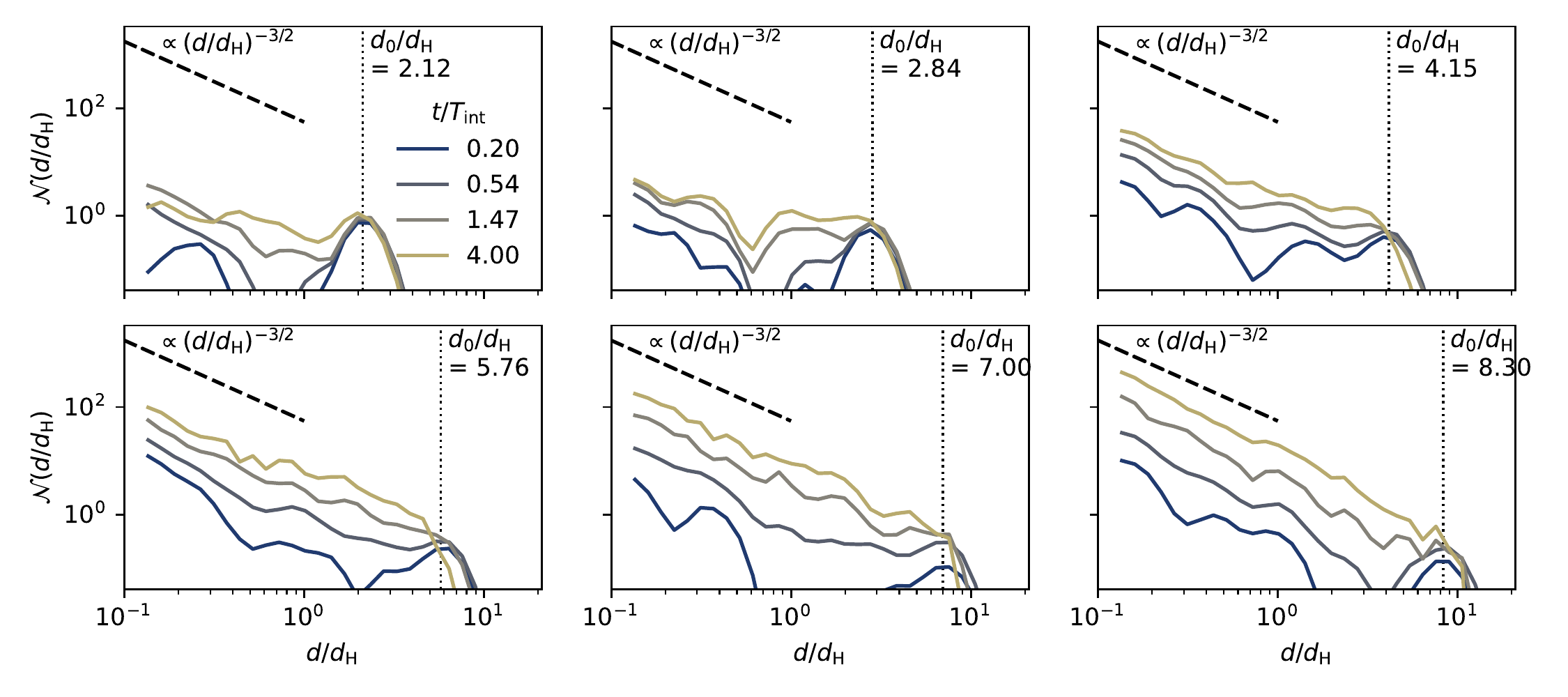}
\put(3,41){(a)}
\put(35,41){(b)}
\put(66.5,41){(c)}
\put(3,22){(d)}
\put(35,22){(e)}
\put(66.5,22){(f)}
\end{overpic}
\caption{Bubble size distributions during the disintegration of cavities with $\dodH$ between 2.1 and 8.3 and times up to $4 T_\mathrm{int}$ after the cavity is released into the turbulence. The size of the parent bubble is denoted by the dashed vertical line. Each distribution integrates to the average number of bubble observed at that condition at that time. The eventual sub-Hinze power-law scaling exponent approaches $-3/2$, shown by the \dan{dashed black} line, as $d_0/d_\mathrm{H}$ is increased.}
\label{fig:bubblerelease_transient_distributions_SDS0}
\end{figure}

The experiment was carried out with six values of $d_0 / d_\mathrm{H}$ between 2.1 and 8.3, with 10---20 runs taken at each condition, as given in \Cref{tab:cavity_conditions}. \dan{Note that the largest cavities exceed the integral length scale in size, so the typical turbulent stress at their spatial scale will be saturated relative to that predicted by the Kolmogorov scaling employed in the definition of the Hinze scale.} \Cref{fig:bubblerelease_transient_distributions_SDS0} shows the transient evolution of $\mathcal{N}(d/d_\mathrm{H}) = N(d) \dH$ for each condition (ensemble-averaging the 10---20 runs). At early times, the distributions for all $\dodH$ exhibit a peak at $\dodH$, denoted by the vertical dotted lines. For the two smallest cavities (given in \Cref{fig:bubblerelease_transient_distributions_SDS0} panels (a-b)), for which no break-up was observed during many runs, only a small number of bubbles are formed over time, and the size distribution near the injection scale does not decrease appreciably with time.

Over time, as the larger cavities (given in \Cref{fig:bubblerelease_transient_distributions_SDS0} panels (c-f)) begin to disintegrate, the size distribution for $d<d_0$ begins to be built up. Even among these larger cavities which produce a considerable number of sub-Hinze bubbles, the increase in the number of sub-Hinze bubbles is much more pronounced for the cavities that are initially larger (evidenced by comparing curves for $\dodH=4.15$ and $\dodH = 8.30$, for example).  This suggests that there is a large separation of scales between the sub-Hinze bubbles and the parent bubbles responsible for their creation; more simply, large bubbles are needed for the production of small bubbles. For the largest cavities, the size distribution for sub-Hinze bubbles eventually follows an $\mathcal{N}(\ddH) \propto (\ddH)^{\alpha_d}$ scaling, with $\alpha_d = -3/2$, sketched on all plots as the dashed line for reference. The final curves shown (for $t/T_\mathrm{int}=4$) might constitute an under-estimation for the bubble size distribution for smaller bubbles, since some of the bubbles which may break have risen out of the field of view by this time. 

\begin{figure}
\centering
    \begin{overpic}[width=0.9\linewidth]{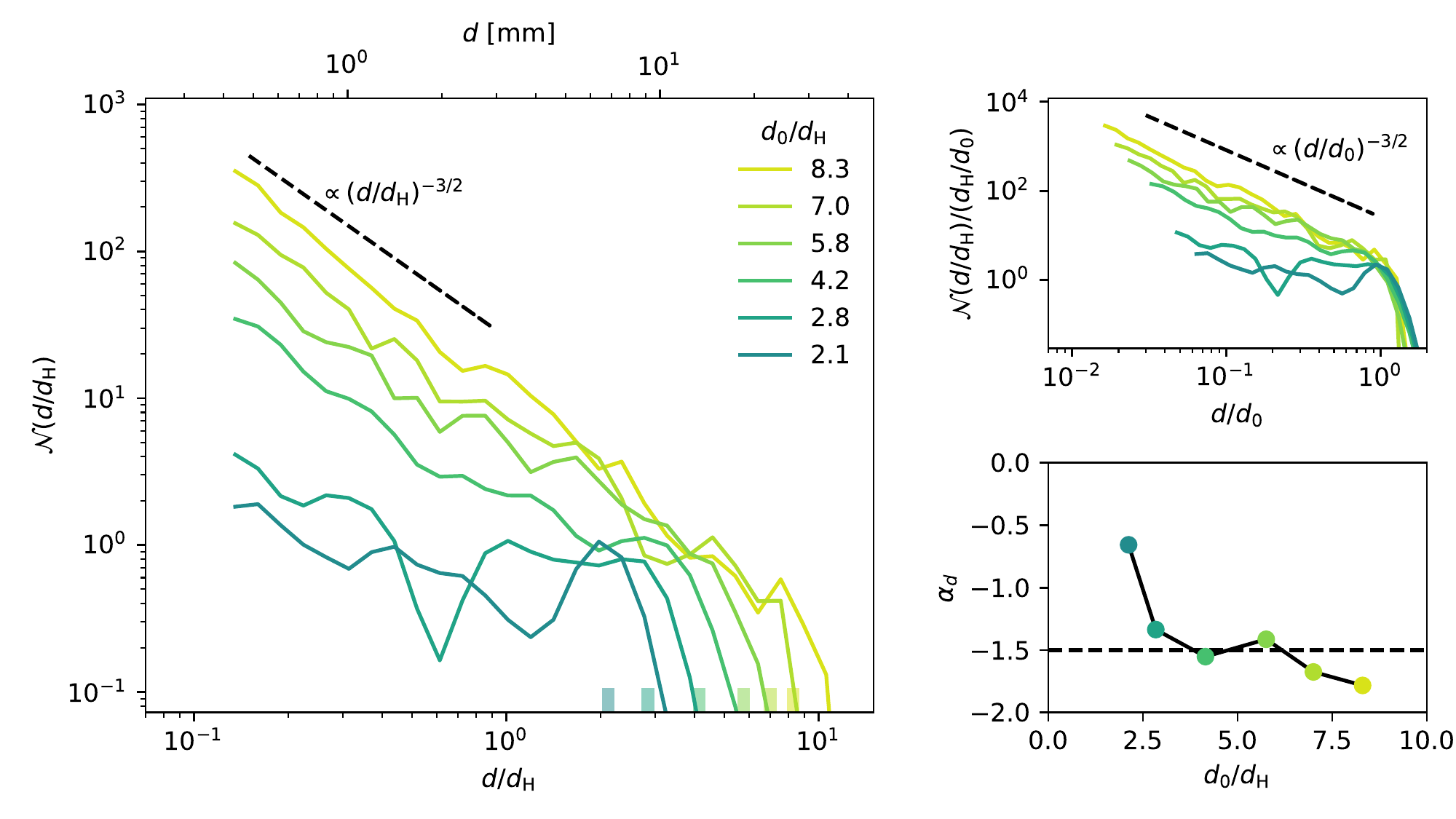}
    \put(10.5,46.5){(a)}
    \put(72,46.5){(b)}
    \put(72,21.5){(c)}
    \end{overpic}
\caption{Time-averaged bubble size distributions. (a) The \changed{dimensionless} bubble size distribution averaged between $t/T_\mathrm{int}=2$ and $t/T_\mathrm{int}=4$ for cases with varying $\dodH$, denoted by the position of the colored notches along the bottom axis. Distributions for $\dodH<2$ are smoothed slightly to account for the small number of observations. \changed{(b) The bubble size distributions based on the diameter normalized by the initial cavity diameter $d_0$.} (c) The exponent $\alpha_d$ of a power-law fit to the sub-Hinze portion of each distribution, $\mathcal{N}(\ddH) \propto (\ddH)^{\alpha_d}$ for $\ddH < 1$, indicating that as $\dodH$ is increased, the sub-Hinze spectrum approaches a $\mathcal{N}(\ddH) \propto (\ddH)^{-3/2}$ scaling.}
\label{fig:bubblerelease_final_size_distributions}
\end{figure}

\changed{Now, we consider the size distributions averaged averaged between $2 T_\mathrm{int}$ and $4 T_\mathrm{int}$. During these times, a significant number of break-ups have occured (for larger $\dodH$), but a significant portion of the bubbles have not yet left the field of view, and the bubble size distribution approaches a constant shape.} \Cref{fig:bubblerelease_final_size_distributions} (a) compares the size distributions over these times for each value of $\dodH$. For larger air cavities, the magnitude of $\mathcal{N}(\ddH)$ is increased, and the sub-Hinze power-law distribution steepens. \changed{The same data is shown in panel (b), normalized by the cavity diameter $d_0$ instead of the Hinze scale. Larger cavity sizes yield a $\propto d^{-3/2}$ scaling for all bubble sizes.} 

\Cref{fig:bubblerelease_final_size_distributions} (c) shows the power-law exponent fit to the sub-Hinze portion of the distributions in panel (a), $\mathcal{N}(\ddH) \propto (\ddH)^{\alpha_d}$ for $\ddH < 1$, for each case. As $\dodH$ is increased, \dan{an $\alpha_d = -3/2$} scaling is approached, indicated by the dashed black line. The size distribution is affected not only by the break-up physics, but is also steepened by the rising dynamics of the bubbles: as small bubbles rise more slowly than larger ones, they tend to linger in the field of view for longer, increasing their concentrations \citep{Garrett2000}.

\begin{figure}
\centering
  \begin{overpic}[width=1\linewidth]{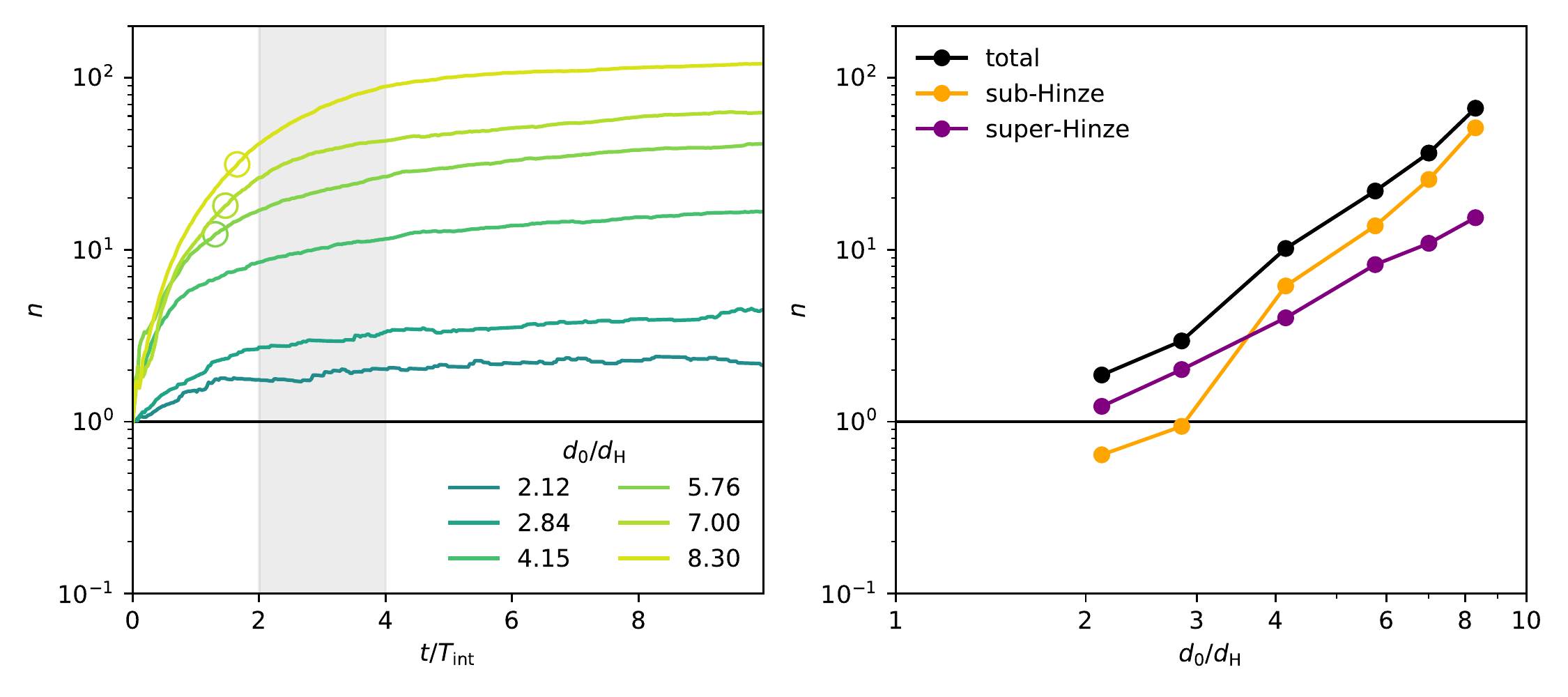}
  \put(3,42){(a)}
  \put(52,42){(b)}
  \end{overpic}
\caption{Evolution of the number of resolved bubbles (limited to $d/\dH > 0.12$) with time and the initial cavity size. (a) Temporal evolution of the average number of all bubbles measured experimentally for different initial cavity sizes $\dodH$. Circles give the values employed in \cref{sec:model_m}. (b) The total number of bubbles (black), number of sub-Hinze bubbles (orange), and number of super-Hinze bubbles (purple) averaged between $2 T_\mathrm{int} \leq t < 4 T_\mathrm{int}$ (the region shaded in gray in (a)) as a function of the initial cavity size.}
\label{fig:num_vs_time_and_d0dH}
\end{figure}

Integrating the transient size distributions over the bubble diameter, the temporal evolution of the number of resolved bubbles \changed{$n$ (with the minimum resolvable size $d_\mathrm{min}=0.12 \dH$)} is shown in \Cref{fig:num_vs_time_and_d0dH} (a). The gray shaded region denotes the times over which the bubble size distributions are averaged in \Cref{fig:bubblerelease_final_size_distributions} and \Cref{fig:num_vs_time_and_d0dH} (b).

\Cref{fig:num_vs_time_and_d0dH} (b) shows the number of resolved sub-Hinze, super-Hinze, and total bubbles, averaged over the time period considered, with the minimum resolved bubble size $d_\mathrm{min} \approx 0.12 \dH$. Again, we see an increase in the number of bubbles formed with the initial size of the cavity. Further, the number of sub-Hinze bubbles increases with the parent bubble size more rapidly than the number of super-Hinze bubbles does, making sub-Hinze bubbles constitute a larger portion of the bubble size spectrum for larger $d_0/\dH$. This is remarkable, since as $d_0/\dH$ is increased, the span of bubble sizes constituting resolvable sub-Hinze bubbles ($d_\mathrm{min} < d < \dH$) remains fixed, while the span of potential super-Hinze bubble sizes ($\dH < d < d_0$) increases.

Taken together, \Cref{fig:bubblerelease_final_size_distributions,fig:num_vs_time_and_d0dH} are congruent with the capillary pinching mechanisms for sub-Hinze bubble production put forward by \cite{Riviere2021cap}. Our figures suggest that their formation relies on the break-up of cavities that are significantly larger than the Hinze scale: only larger values of $d_0/d_\mathrm{H}$ yield the $\mathcal{N}(d/\dH) \propto (d/\dH)^{-3/2}$ power-law scaling in the sub-Hinze bubble size distribution, and the dependence on $d_0$ of the number of sub-Hinze bubbles produced (shown in \Cref{fig:num_vs_time_and_d0dH} (b)) is steeper than that of the number of super-Hinze bubbles produced. \dan{We propose in the next section an explanation of the mechanisms leading to this  dependence.}

\subsection{Capillary splitting of ligaments prepared by the turbulence produces small bubbles}
\label{sec:concurrent_mechanisms}

Visual observations of the large air cavities disintegrating provide clues into the mechanism responsible for the production of sub-Hinze bubbles: child bubbles much smaller than the Hinze scale are seen to originate from a Rayleigh-Plateau-like instability that occurs during the pinching apart of elongated fluid ligaments prepared by the turbulence. However, the turbulence is only able to deform bubbles that are large enough with respect to the Hinze scale that such ligaments might be created, since surface tension is effective at limiting the severity of deformations to smaller bubbles. These experimental observations parallel \dan{a} recent interpretation of DNSs of bubble break-up \citep{Riviere2021cap}.

\begin{figure}
\centering
  \begin{overpic}[width=0.7\linewidth]{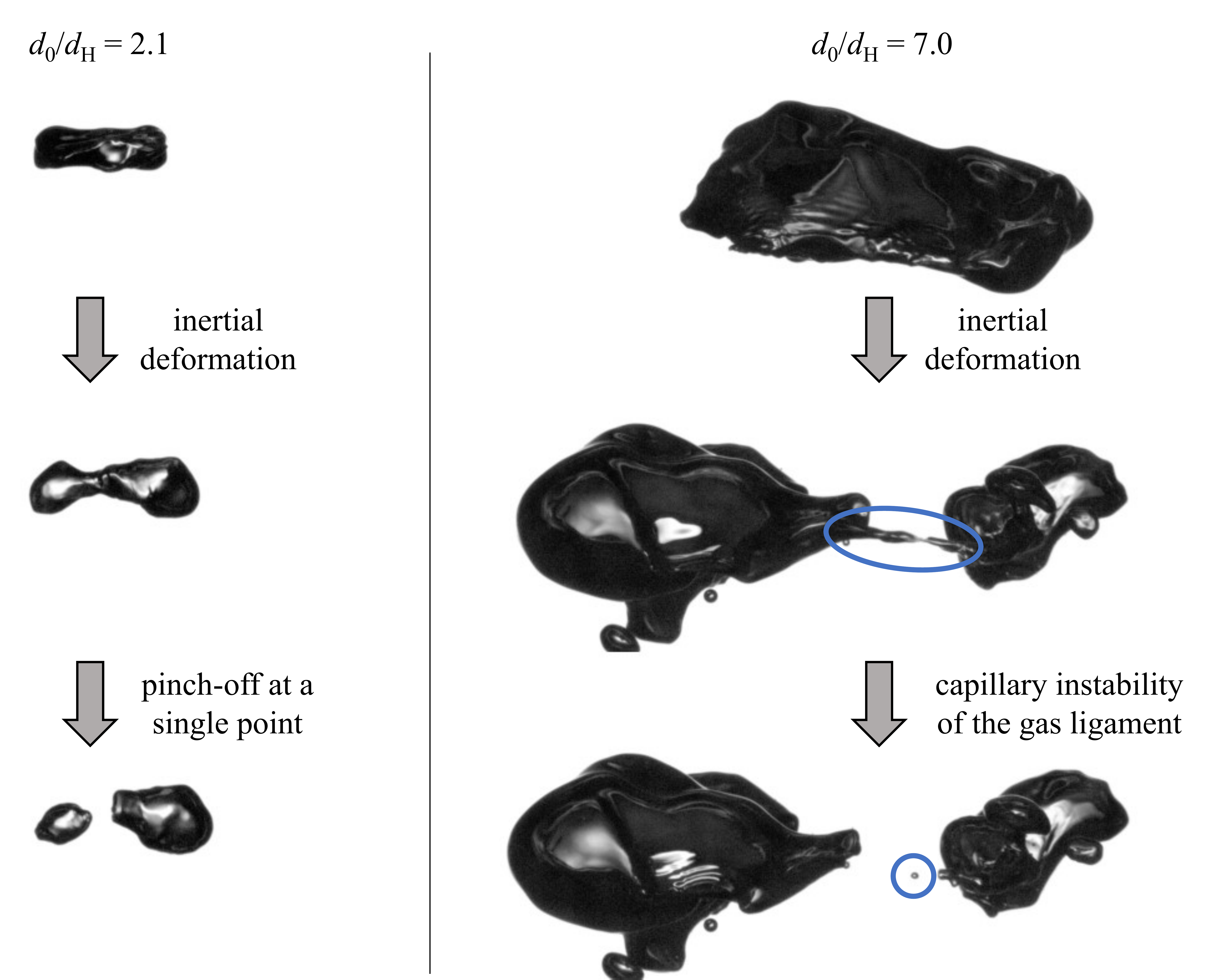}
  \put(-5,75){(a)}
  \put(40,75){(b)}
  \end{overpic}
\caption{Break-up of a bubble initially close to the Hinze scale in size (a, with $d_0/d_\mathrm{H} = 2.1$) and initially much bigger than the Hinze scale (b, with $d_0/d_\mathrm{H} = 7.0$). The deformation to the smaller bubble produces two comparably-sized lobes, which split apart to form two comparably-sized child bubbles. The deformation to the larger bubble also produces two comparably-sized lobes, but these are separated by a much more elongated filament of air. The unstable collapse of this filament produces the small "capillary" child bubble between the two larger ones. \changed{The small bubbles in the lower left of the image were produced in previous break-ups.}}
\label{fig:explanation_figure_small_fig}
\end{figure}

Illustrative examples of bubble break-up are given in \Cref{fig:explanation_figure_small_fig}, which shows the typical break-ups of bubbles of two sizes: one is near the Hinze scale in size (a), and another is seven times larger than the Hinze scale (b). The smaller bubble, with $d_0/d_\mathrm{H} = 2.1$, is initially deformed into two comparably-sized lobes, and the neck separating the two splits at a single point to form two child bubbles, each of a similar scale as the parent. Here, the parent bubble is small enough that surface tension is able to prevent significant deformation during the break-up.

The larger bubble, with $d_0/d_\mathrm{H}=7.0$, is similarly deformed by the turbulence into two comparably-sized lobes prior to pinch-off. However, the filament of air separating the two just prior to pinch-off has become significantly more elongated than the neck in the break-up of the smaller bubble. This elongation opens the door to capillary instabilities along the filament during its collapse: in the instance shown in \Cref{fig:explanation_figure_small_fig} (b), the filament pinches apart at two separate points, leaving a small child bubble (with $d \ll d_\mathrm{H}$) between the two lobes.

\dan{The two examples of break-up discussed illustrate} two mechanisms present in the break-up of bubbles by turbulence. The first is the deformation of the parent bubble by a turbulent structure, likely on the spatial scale of the parent bubble itself. This brings the bubble to an unstable state consisting of two lobes (which will become what we call the "inertial" child bubbles) separated by a neck of air, which begins to pinch apart under capillarity. When the deformation to the bubble is severe enough, this ligament can take on an elongated, deformed shape. Its pinching can become unstable under a Rayleigh-Plateau-like mechanism, leading to the formation of small "capillary" bubbles.

\begin{figure}
\centering
  \begin{overpic}[width=1\linewidth]{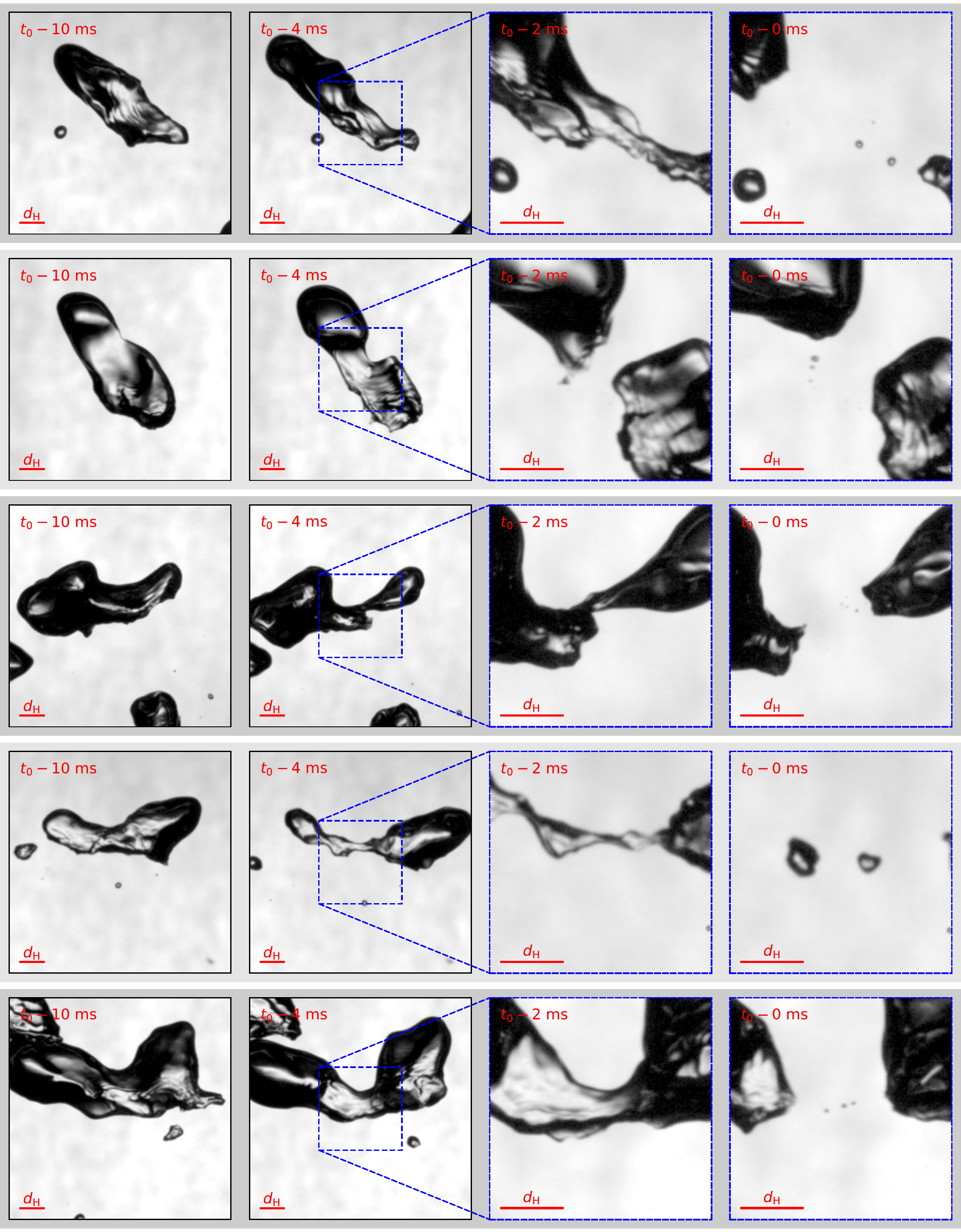}
  \end{overpic}
\caption{\changed{Five cases of sub-Hinze bubble production by the unstable collapse of a deformed ligament. Each row shows four snapshots in time, spaced 10, 4, 2, and \SI{0}{ms} before the time at which the sub-Hinze bubbles are first visible. The field of view in the final two columns is given by the blue square in the second column.}}
\label{fig:instability_cases}
\end{figure}

\changed{\Cref{fig:instability_cases} shows an additional five instances of deformed ligaments undergoing a capillary instability to produce sub-Hinze bubbles. Many cases, especially those involving large parent bubbles, do not solely involve one ligament separated by two lobes; turbulent deformations cause the bubble shapes to be more irregular. However, in all instances, very small bubbles are produced as an air ligament involved in the turbulent deformation collapses unstably.}

This \changed{description} is clearly a simplified understanding of the bubble pinch-off process, as it does not capture the redistribution of air due to a capillary pressure difference between lobes that may be responsible for the formation of small bubbles \citep{Andersson2006}, nor does it describe the "tearing off" of very small bubbles that we \changed{observe occurring to large parent bubbles more frequently than $1/T_\mathrm{turb}(d_0)$}. However, the framework serves as a bridge between the inertial deformations to a bubble by the turbulence and the later-time collapse dynamics instigated by capillarity. This understanding mirrors the description of bubble pinch-off in turbulence given in \cite{Ruth2019}, in which we showed that turbulence \dan{sets} an "initial" deformed bubble shape before the collapse dynamics overtake the turbulent dynamics. Once the inertial collapse of the neck becomes fast enough (equivalently, once the neck becomes small enough), however, the turbulence effectively "freezes" in place relative to the accelerating collapse dynamics. The end result is that the final stage of the pinching process---in this case, the production of small bubbles through the capillary instability of gas ligaments---is affected by the turbulence only insofar as the turbulence sets the "initial condition" on which the remainder of the process evolves under capillary and, eventually, inertial, dynamics.

\FloatBarrier
\section{Individual break-up event dynamics}
\label{sec:dynamical}

So far, we have considered the transient size distributions $\mathcal{N}(\ddH)$ that result from air cavities with $d_0 > d_\mathrm{H}$ being introduced to turbulence. In this section, we focus on individual break-up events that are tracked individually in three dimensions as described in \Cref{sec:exp_dynamical}; these events are the building blocks for the disintegration of larger cavities.

\dan{We will characterize the break-up events over their typical time scale, which is given by the eddy turn-over time at the parent bubble's scale, $T_\mathrm{turb}(d_0) = \epsilon^{-1/3} d_0^{2/3}$, following discussions from \citet{Risso1998,Martinez-Bazan1999a,Riviere2021jfm}.}

\subsection{Qualitative discussion of the break-up sequences}

\begin{figure}
\centering
  \begin{overpic}[width=0.9\linewidth]{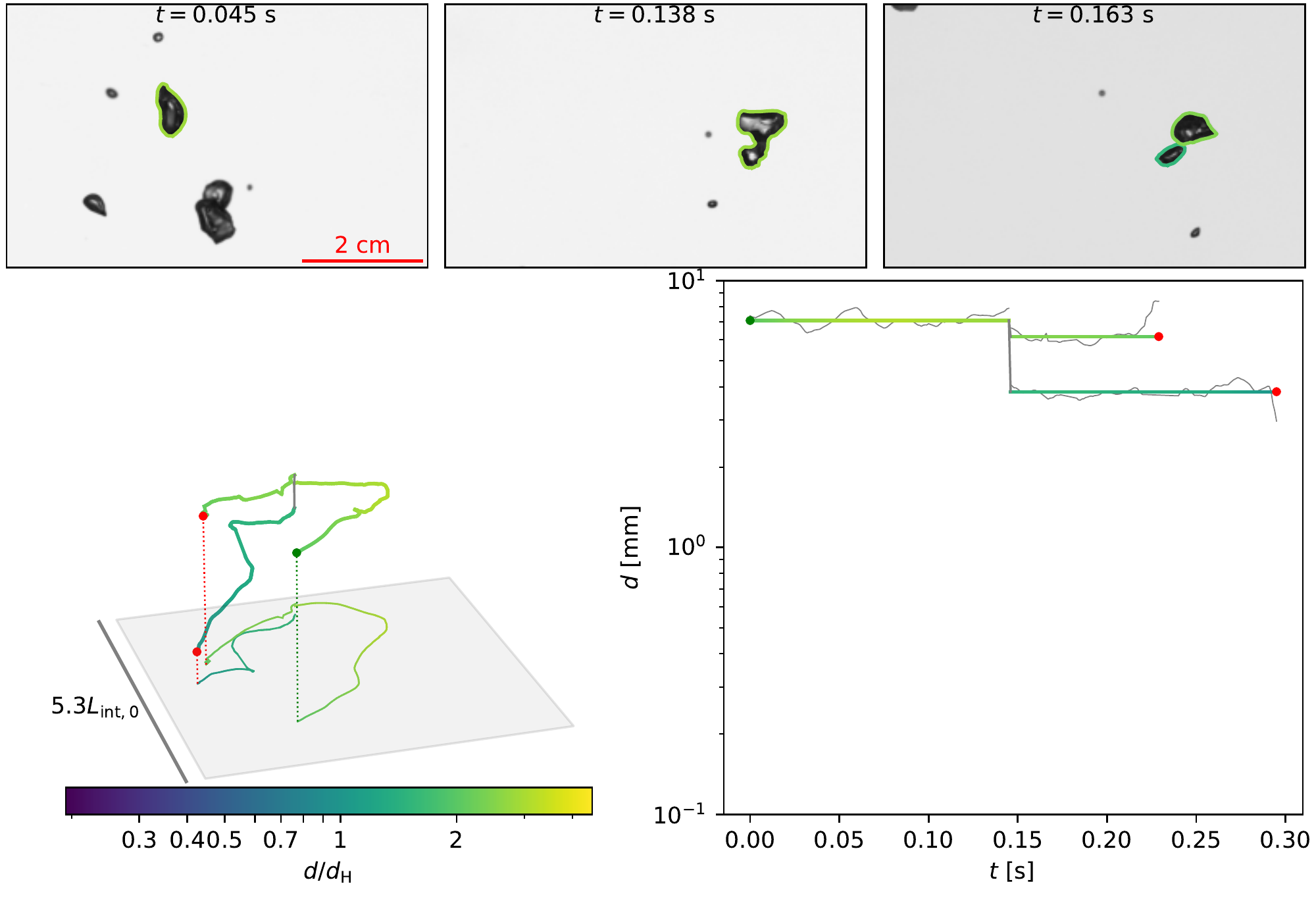}
  \put(1,65){(a)}
  \put(34,65){(b)}
  \put(67,65){(c)}
  \put(1,30){(d)}
  \put(47,45){(e)}
  \end{overpic}
\caption{One dynamically-tracked bubble break-up involving the production of $m=2$ child bubbles. (a-c) Images recorded by one of the two high-speed cameras throughout the sequence. (d) The trajectories taken by the bubbles involved, with their size relative to the Hinze scale encoded in the color. The green circle marks the first observation of the parent bubble, and the red circles mark the final observation of the child bubbles. The side length of the square shown is given in terms of the integral length scale at the break-up location. (e) The "family tree" for the single break-up, giving diameters of the bubbles present at each point in time. Fainter lines give the instantaneously-measured diameters, and straight lines give the median for each bubble, which is the quantity we consider in our analysis.}
\label{fig:dynamical_intro_figure_binary}
\end{figure}

\begin{figure}
\centering
  \begin{overpic}[width=0.9\linewidth]{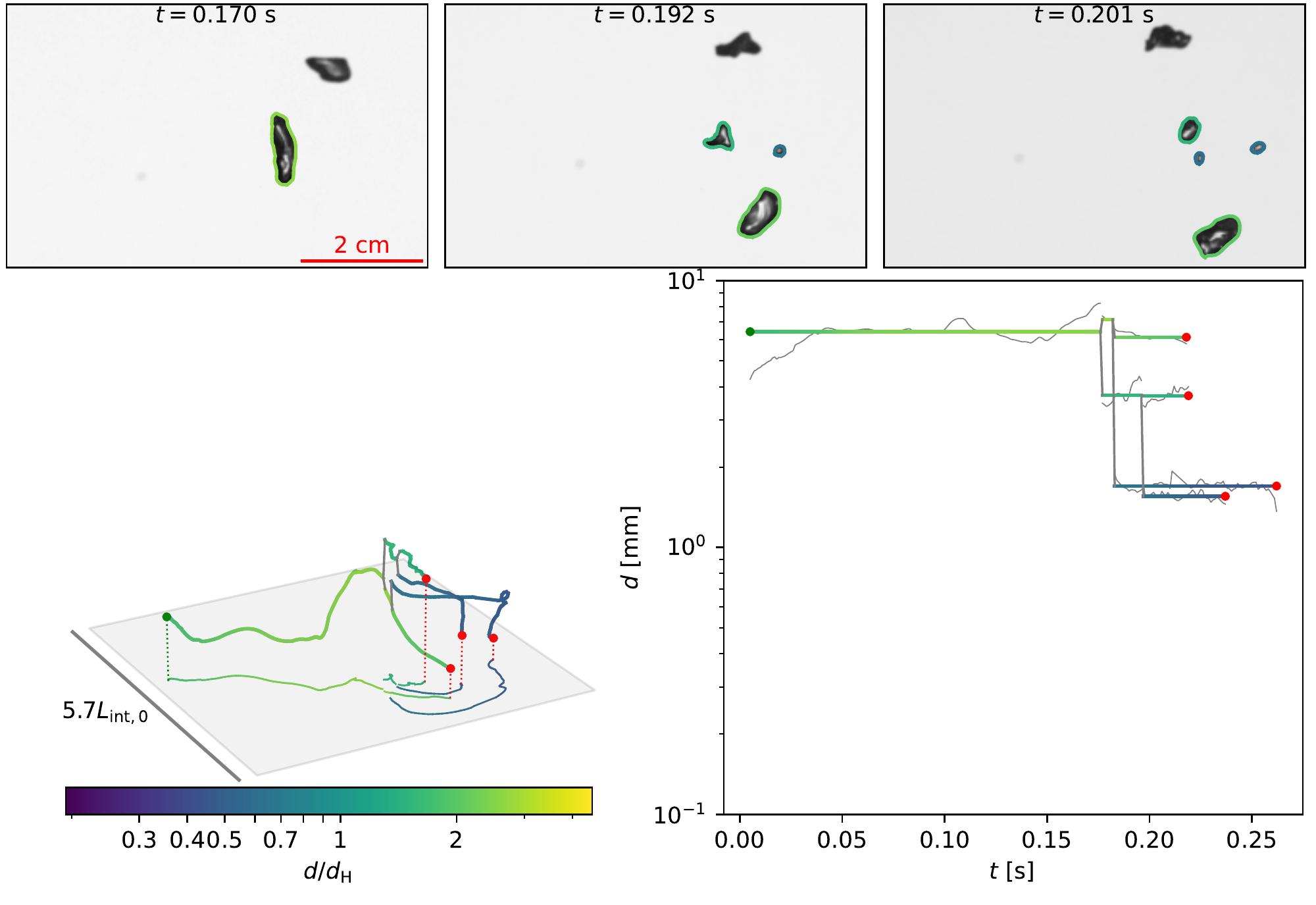}
    \put(1,65){(a)}
  \put(34,65){(b)}
  \put(67,65){(c)}
  \put(1,30){(d)}
  \put(47,45){(e)}
  \end{overpic}
\caption{One dynamically-tracked bubble break-up involving the production of $m=4$ child bubbles. (a-c) Images recorded by one of the two high-speed cameras throughout the sequence. (d) The trajectories taken by the bubbles involved, with their size relative to the Hinze scale encoded in the color. The green circle marks the first observation of the parent bubble, and the red circles mark the final observation of the child bubbles. The side length of the square shown is given in terms of the integral length scale at the break-up location. (e) The "family tree" for the single break-up, giving diameters of the bubbles present at each point in time. Fainter lines give the instantaneously-measured diameters, and straight lines give the median for each bubble, which is the quantity we consider in our analysis.}
\label{fig:dynamical_intro_figure_multiple}
\end{figure}

One break-up producing $m=2$ child bubbles is shown in \Cref{fig:dynamical_intro_figure_binary}, and one producing $m=4$ bubbles is shown in \Cref{fig:dynamical_intro_figure_multiple}. In each, images throughout the break-up sequence are shown in (a-c), and the three-dimensional trajectories taken by the bubbles involved are shown in (d). At each point, the bubble's size is computed relative to the local Hinze scale, and $d/\dH$ is encoded in the trajectory color. The spatial scale is given in terms of the integral length scale at the break-up location, $L_{\mathrm{int},0}$, showing that the bubble trajectories are resolved over multiple integral length scales. Panel (e) shows the dimensional diameters of the bubbles involved over time. 

In the case of binary break-up, given in \Cref{fig:dynamical_intro_figure_binary}, the parent bubble enters the imaged volume from the foreground, and quickly encounters a region of more intense turbulence, where $\dodH$ increases. Eventually, having become deformed, the bubble pinches apart into two child bubbles, each of which are comparable in size to the parent. Both child bubbles persist in the field of view for at least a tenth of a second ($\sim$ an integral-scale turn-over time) without breaking.

In the more complex break-up shown in \Cref{fig:dynamical_intro_figure_multiple}, the parent bubble similarly traverses from a region of less intense turbulence to more intense turbulence, increasing the value of $\dodH$. Eventually, at $t=\SI{0.170}{s}$ (shown in panel (a)), the bubble becomes elongated in the vertical direction, and in a sequence of two rapid splitting events produces the three child bubbles that are visible at $t=\SI{0.192}{s}$ (shown in panel (b)). One is still larger than $\dH$, one is of the order of $\dH$, and the third, left between the two, is smaller than $\dH$. The bubble of the order of the Hinze scale is still significantly deformed, the capillary dynamics involved with the break-up not yet having relaxed. A short time later, by $t=\SI{0.201}{s}$ (shown in panel (c)), an additional bubble has split from it, leaving a total of four child bubbles.

\subsection{Identification of break-up events}
\label{sec:dynamical_identification}

We identify bubble break-ups like the ones shown in \Cref{fig:dynamical_intro_figure_binary,fig:dynamical_intro_figure_multiple} as being sequences of bubble splitting events not exceeding the eddy turn-over time at the parent bubble scale, $T_\mathrm{turb}(d_0) = \epsilon^{-1/3} d_0^{2/3}$. To enforce this temporal constraint, we first construct a "family tree" of all splitting events recorded in one run. Then, if any bubble is present at a time $T_\mathrm{turb}(d_0) $ beyond the initial detected break-up of the first bubble (with diameter $d_0$), we truncate the family tree at that bubble, and start a new family tree with the same bubble (if it later breaks apart). After doing so, we store the sizes of the parent bubble and child bubbles, as well as the turbulence characteristics spatially interpolated at the initial break-up location. To remove spurious break-ups, we discard those for which the sum of the calculated volumes of the $m$ child bubbles is less than 50\% of, or more than 200\% of, the calculated volume of the parent bubble.

\begin{figure}
\centering
  \begin{overpic}[width=1\linewidth]{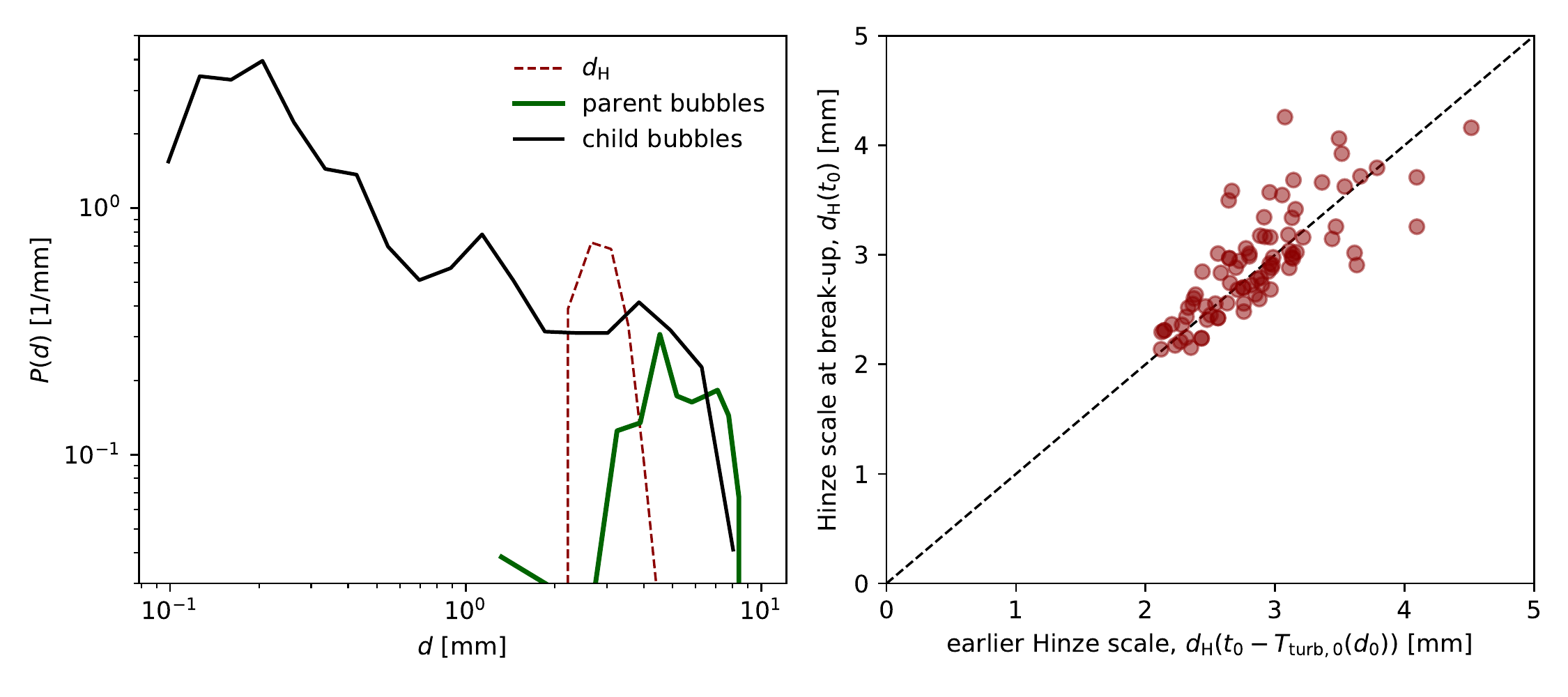}
    \put(9,40){(a)}
  \put(56.5,40){(b)}
  \end{overpic}
\caption{Results on individual bubble break-ups. (a) Distributions of the Hinze scale at the break-up location (red), parent bubble sizes (green), and of the child bubble sizes (black). \changed{(b) The Hinze scale at the bubble's break-up position (vertical axis) as a function of the Hinze scale at the bubble's position one bubble-scale turn-over time prior to break-up (horizontal axis), for the 52\% of cases in which the bubble was in the volume resolved with PIV at this time.}}
\label{fig:dynamical_dim_nondim_sizedists}
\end{figure}

In total, we captured 162 bubble break-ups with this dynamical tracking approach that fit the volume conservation criteria. \Cref{fig:dynamical_dim_nondim_sizedists} (a) shows the distributions of the break-up conditions (the Hinze scale at the break-up location and the parent bubble size) for the aggregated dataset, which we later break down by the parent bubble's size relative to the Hinze scale. The parent bubble diameter $d_0$ is typically slightly larger than the Hinze scale, as the distribution of $d_0$ \changed{(the green line)} is located just to the \changed{right} of that of $d_\mathrm{H}$ (the \changed{dashed red} line). Thus, the break-ups we capture in this experiment have $d_0/d_\mathrm{H} \approx 0.4 \text{--} 3.7$. The black curve shows the distribution of the sizes of child bubbles formed during break-ups, integrating to the average number of bubbles formed per break-up event. 

\changed{To gauge the effect of inhomogeneity in the generated turbulence, we consider how the local turbulence intensity experienced by the bubble (in a Lagrangian sense) varies over timescales relevant to the bubble's break-up. Ideally, a bubble would not be advected through statistically inhomogeneous turbulence during the course of its break-up. Denoting the Hinze scale at the bubble's location at time $t$ as $d_\mathrm{H}(t)$, \Cref{fig:dynamical_dim_nondim_sizedists} (b) shows the Hinze scale at the break-up location $\dH(t_0)$ as a function of the Hinze scale at the bubble's location one bubble-scale eddy turn-over time prior, $\dH(t_0 - T_\mathrm{turb}(d_0))$ for the 52\% of observed break-ups in which the bubble is inside the volume resolved by PIV (in which we are able to compute $\dH$) at this point in time. There is little difference between $\dH(t_0)$ and $\dH(t_0 - T_\mathrm{turb}(d_0))$, suggesting that the local turbulence characteristics experienced by the bubble do not change appreciably during the break-up, and that the turbulence is homogeneous over scales relevant to the break-up.}

\begin{figure}
\centering
\begin{overpic}[width=0.889\linewidth]{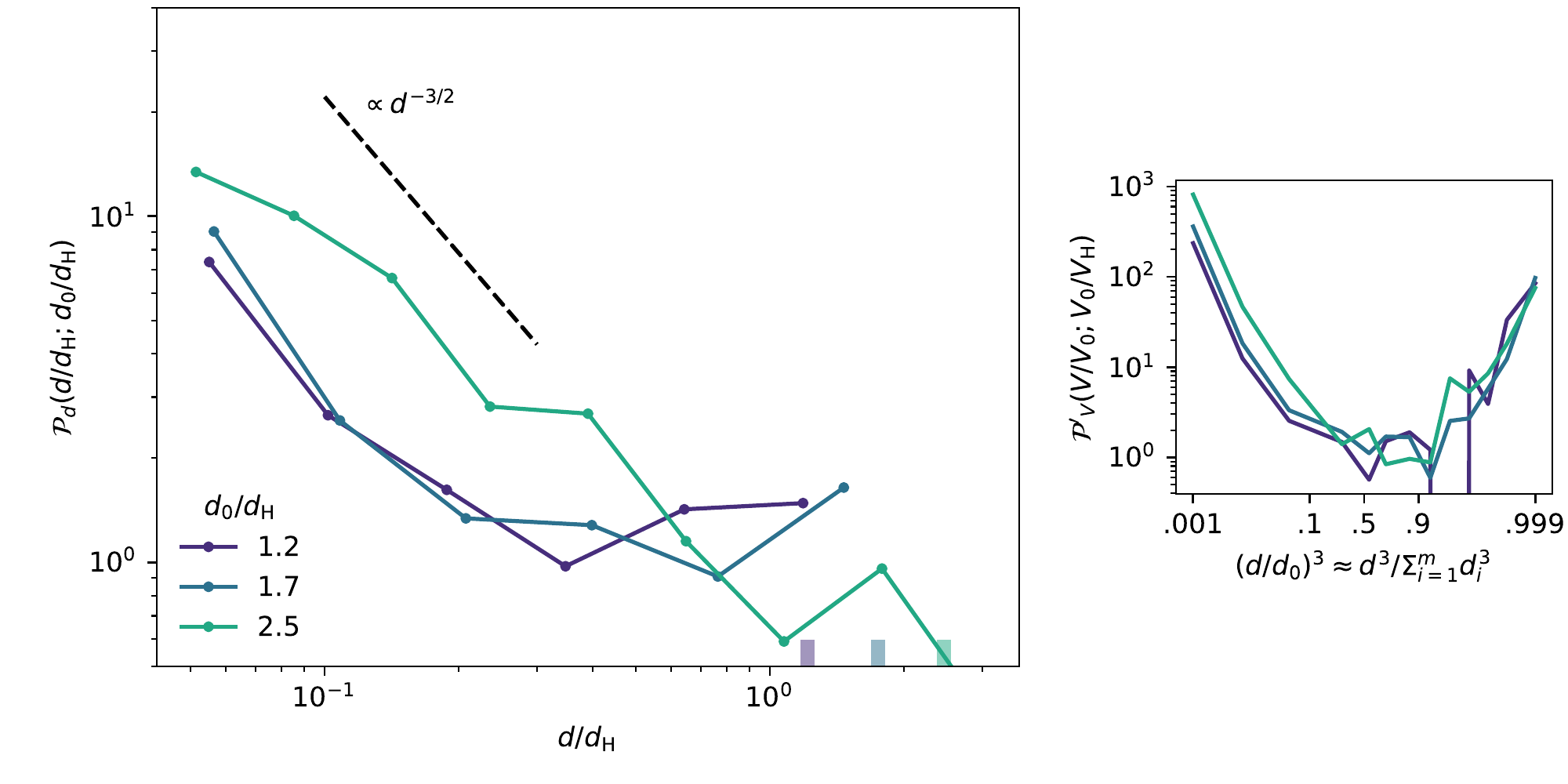}
\put(4,47){(a)}
\put(67,38){(b)}
\end{overpic}
\caption{Dimensionless bubble break-up child size distributions for various approximate values of $d_0/d_\mathrm{H}$. The value give for each curve (which is denoted by the notch on the horizontal axis) corresponds to the mean value of $\dodH$ for that curve. (a) The distributions of child bubble diameter normalized by the Hinze scale. (b) The \changed{volumetric child size distribution}, with child bubble volumes normalized by the parent bubble volume, which is approximated as the sum of the resolved child bubble volumes.}
\label{fig:dynamic_nondim_childsizedists}
\end{figure}

\subsection{Child size distribution}

Now, we compute the dimensionless bubble child \changed{size} distributions conditioned on the approximate dimensionless parent bubble size, $\mathcal{P}_d(d/d_\mathrm{H};d_0/d_\mathrm{H})$. The data is averaged over three ranges of $d_0/d_\mathrm{H}$ (the ranges between [0.3:1.55];[1.55:1.93]; and [1.93:3.70]), and results are shown in \changed{\Cref{fig:dynamic_nondim_childsizedists}} (a). As $d_0/\dH$ is increased, the dependence of $\mathcal{P}_d$ on $\ddH$ becomes steeper. The dashed line gives the $\mathcal{P}_d(\ddH;\dodH) \propto (\ddH)^{-3/2}$ scaling, which is approached for large $\dodH$ due to the production of small bubbles by capillary instabilities \citep{Riviere2021cap}. Qualitatively, the child size distribution for smaller parent bubbles is flatter near the Hinze scale, while that for larger parent bubbles increases more rapidly with decreasing bubble size as a power-law relationship. 
\dan{Note that the child size distribution is defined so that it integrates to the average number of child bubbles formed.}

This representation of the child size distribution masks the large number of bubbles formed very close to the parent bubble size. To capture these small bubbles, we also compute the \changed{volumetric child size distribution}, normalized by the volume of the parent bubble $V_0$. Since the determination of the volumes of larger bubbles is difficult given their deformations, we approximate the parent bubble volume as the sum of the volumes of the child bubbles, and consider $(d/d_0)^3 \approx d^3 / \sum_{i=1}^m d_i^3$ \citep{Vejrazka2018}. The distribution of these dimensionless volumes is shown in \Cref{fig:dynamic_nondim_childsizedists} (b), exhibiting a $\cup$ shape that is not strongly dependent on $d_0/\dH$ (though we again see increased small bubble production with larger $d_0/\dH$). The large values of this distribution near 1 suggest that in many break-up events, small bubbles are "torn off" of the parent bubble, without inertial deformation producing multiple child bubbles of sizes comparable to that of the parent. We note that the resolution of our experiment (in which the smallest bubble we can detect is approximately \SI{200}{\micro m} in diameter) limits the number of bubbles detected.


\subsection{Small bubble production without significant inertial deformation}
\label{sec:end_pinching}

\begin{figure}
\centering
  \begin{overpic}[width=0.9\linewidth]{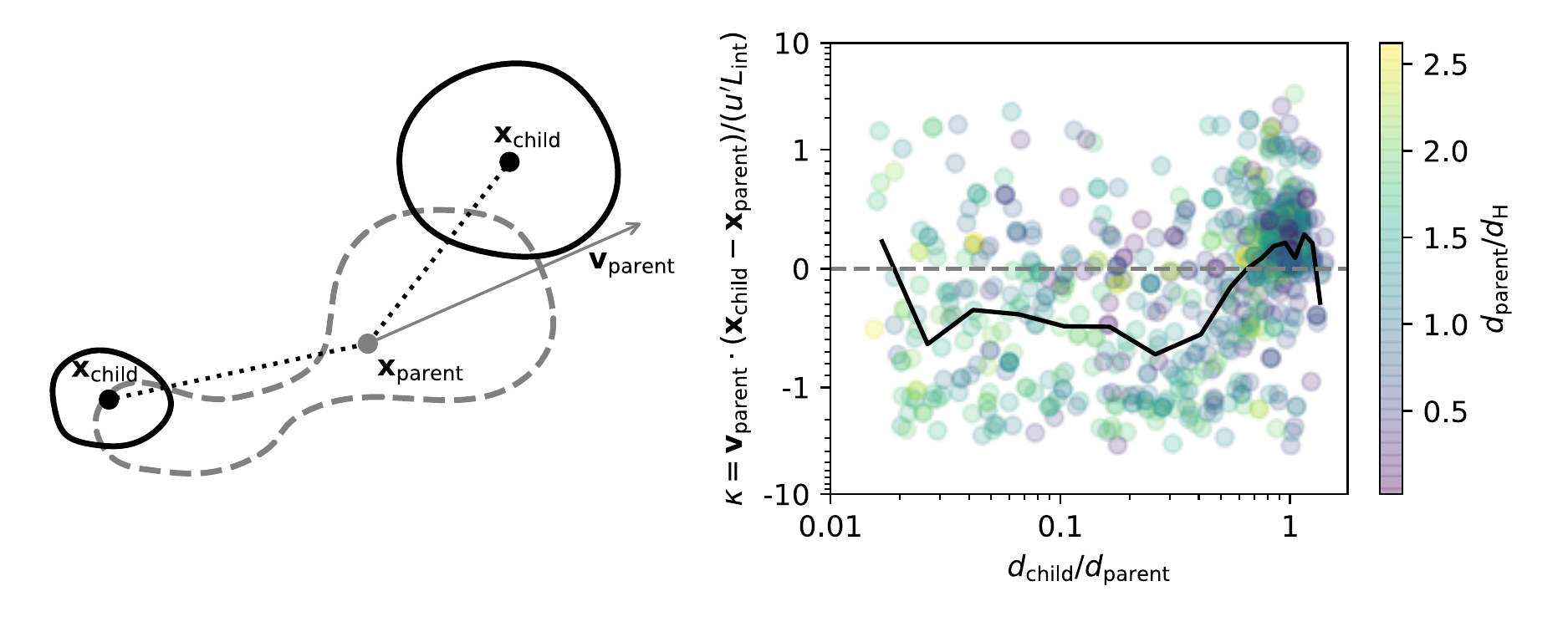}
  \put(1,34){(a)}
  \put(54,34){(b)}
  \end{overpic}
\caption{Statistics of the positions of bubbles after splitting events. (a) A sketch of a splitting event involving small bubble production, including the parent bubble velocity at break-up and the final and initial positions, respectively, of the parent and child bubbles. (b) The initial child bubble position relative to the parent bubble's motion, $\kappa$, for each splitting event (circles), as well as the mean value conditioned on the normalized child bubble size (black line). $\kappa<0$ denotes bubble production behind the parent bubble, while $\kappa>0$ denotes bubble production ahead of the parent bubble.}
\label{fig:dynamic_positions}
\end{figure}

In many of the break-ups we observe in the large cavity disintegration and individual break-up experiments, small bubbles were seen to be "torn off" from a parent bubble, without an appreciable large-scale deformation to the parent bubble. These events are reminiscent of tip-streaming \citep{Montanero2020}. This phenomenon is evidenced by the right side of the $\cup$-shaped child size distributions shown in \Cref{fig:dynamic_nondim_childsizedists} (b), as a child bubble that is nearly the size of the parent is the signature of such break-ups. To understand these events, we present in \Cref{fig:dynamic_positions} a qualitative discussion of the dynamics of individual splitting events. For each splitting event, we compare the velocity of the parent bubble at break-up $\vec{v}_\mathrm{parent}$ (denoted by the gray arrow in panel (a)) to the displacement between the parent bubble's final position $\vec{x}_\mathrm{parent}$ (the gray circle) and the initial positions at which the child bubbles are detected $\vec{x}_\mathrm{child}$ (the black circles). The child bubble's initial detected position ahead of or behind the parent bubble, $\kappa = \vec{v}_\mathrm{parent} \cdot (\vec{x}_\mathrm{child} - \vec{x}_\mathrm{parent}) / (u' L_\mathrm{int})$, normalized by turbulence quantities, is then computed, and is plotted against the child bubble's size relative to the parent size in panel (b). The color of each marker denotes the size of the splitting event's parent bubble relative to the Hinze scale. The black line shows the expected value of $\kappa$ given the normalized child bubble size. Smaller child bubbles (with $d_\mathrm{child}/d_\mathrm{parent} < 0.6$, \dan{below which the mean value of $\kappa$ becomes negative}) tend to be left in the wake of the parent bubble ($\kappa<0$), while larger child bubbles tend to be produced ahead of the parent bubble $(\kappa > 0)$. 

While the conceptual picture for break-up discussed in \Cref{sec:concurrent_mechanisms} describes the role of capillarity during break-ups involving large-scale deformations, it is likely that break-ups solely involving small bubble production are also regulated by capillarity: in these cases, a turbulent motion smaller than the parent bubble may succeed in producing a ligament which extends off of one side of the parent, and this ligament may pinch apart into many small bubbles in a capillary instability as it is retracted back into the bulk of the parent bubble. Specifically, \Cref{fig:dynamic_positions} suggests that the bulk of a bubble may often be swept forward by a turbulent eddy, and the trailing ligament may become unstable as it "catches up" with the rest of the parent bubble. Similar to the framework presented in \Cref{sec:concurrent_mechanisms}, the process is initiated by a turbulent deformation to the parent, and ends with the capillary instability of a ligament involved in the deformation. 

\FloatBarrier
\section{A model for bubble break-up}
\label{sec:model}


\subsection{Physical ideas}
\label{sec:model_timescales}



The experiments presented in \Cref{sec:air_cavity_results,sec:dynamical}, taken together with the existing literature, point to \changed{three important time scales that must be considered in developing a population balance model: the inverse of the break-up frequency, the break-up duration, and the capillary capillary pinching times.}

\changed{The longest of these is the typical duration until a break-up occurs---that is, the inverse of the break-up frequency, $1/\omega(d_0)$. This time scale will control how many break-up events will occur over a given time and will be a function of $\dodH$. The second timescale is that over which a break-up typically occurs, or the event duration (i.e., lasting from the start of the deformation until the child bubbles have all been formed), and will also be a function of $\dodH$. The break-ups taking the longest time will be those instigated by the largest eddies capable of causing break-up, which are taken to be those at the parent bubble's scale \citep{Luo1996}. Thus, an upper bound and typical scale of the break-up duration is taken to be the eddy turn-over time at the parent bubble's scale, $T_\mathrm{turb}(d_0) = \epsilon^{-1/3} d_0^{2/3}$, in agreement with experimental and numerical observations of the time over which bubbles are deformed prior to break-up \citep{Risso1998,Martinez-Bazan1999a,Riviere2021jfm}. 
\dan{The final timescale we consider is that of the capillary instabilities of gas ligaments that produce a small child bubble of size $d$, which will occur over the capillary timescale of that child bubble, $T_\mathrm{cap}(d) = (\rho/\gamma)^{1/2} d^{3/2} / (2 \sqrt{3})$ \citep{Riviere2021cap}.}}

\dan{From these three relevant time scales, we define three types of events. At the shortest time, we define the individual binary \textit{splitting events}. For the production of bubbles with $d \ll \dH$, we have $T_\mathrm{cap}(d) \ll T_\mathrm{turb}(d_0)$. At the eddy turn-over time, we define a \textit{break-up} as being a sequence composed of all the splitting events occurring in a time bounded by \changed{$\Delta T_\mathrm{break-up} = T_\mathrm{turb}(d_0)$}, which permits the production of more than two bubbles in a single event (similar to the definition used for drop break-ups by \cite{Solsvik2016}).} Finally, following the nomenclature from \cite{Hinze1955}, a \textit{disintegration} is a longer-duration process involving an arbitrary number of break-ups.

\begin{figure}
\centering
  \includegraphics[width=0.8\linewidth]{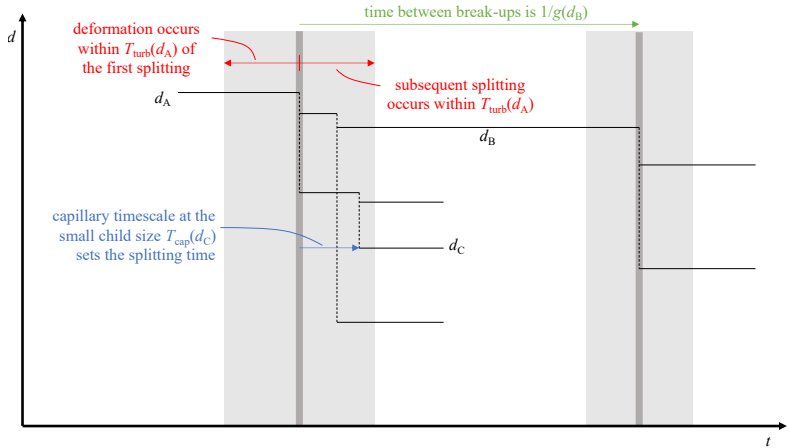}
\caption{Sketch of two bubble break-ups and the associated timescales, with $T_\mathrm{turb}(d) = \epsilon^{-1/3}d^{2/3}$ and $T_\mathrm{cap}(d) = (\rho/\gamma)^{1/2} d^{3/2} / (2 \sqrt{3})$. The gray vertical lines denote the times associated with each of the two break-ups. The shaded region to the left bounds the time over which the deformation to the parent bubble is assumed to occur (the turbulent timescale at the parent bubble size), and the region to the right of the line bounds the time over which the subsequent splitting events are assumed to occur (also taken to be the same turbulent timescale). During the subsequent splitting events, the capillary timescale at the size of the smaller child bubble sets the time over which the splitting event occurs \citep{Riviere2021cap}. The time between break-ups is set by the inverse of the break-up frequency $\omega$ of the bubble which is to break, which we address later in the paper.}
\label{fig:family_tree_timescales}
\end{figure}

These timescales are sketched in \Cref{fig:family_tree_timescales}, which illustrates two break-up events that stem from a bubble of diameter $d_\mathrm{A}$ encountering turbulence. The deformation to the parent bubble that instigates the break-up is assumed to happen within a time $T_\mathrm{turb}(d_\mathrm{A})$ before the first bubble splits from the parent. Then, within an additional time bounded by $T_\mathrm{turb}(d_\mathrm{A})$, subsequent splitting events occur due to capillary instabilities arising from the deformation. One such instability produces a bubble with diameter $d_\mathrm{C}$, and the time over which this instability develops is set by the capillary timescale at the smaller child bubble size, $T_\mathrm{cap}(d_\mathrm{C})$. Later on, one of the child bubbles produced in the first break-up, with diameter $d_\mathrm{B}$, itself breaks up. 

Using these ideas, we propose a population balance model that integrates these physical elements and models the evolution of a bubble size distribution with a Boltzmann transport equation using the bubble size as an internal coordinate. \dan{The population balance model considers a break-up rate kernel $f$, constructed from child size distributions computed through a Monte Carlo approach (constrained by results from experiments and DNSs, informing the number of children and the shape of the distribution) and a parent bubble break-up frequency taken from the literature. With the kernel defined, we integrate the model in time to simulate the evolution of the size distribution during a cavity disintegration and compare to our experimental data.}

\subsection{Population balance modeling}
\label{sec:population_balance_modeling}

In a confined region of homogeneous turbulence, the transient evolution of the absolute dimensionless \changed{volumetric bubble size distribution} $\mathcal{N}_V(\nd{V}) = N_V(V) V_\mathrm{H} = \mathcal{N}(\ddH) / (3 (\ddH)^2)$, where $N_V(V)$ is the absolute dimensional \changed{volumetric size distribution}, $\nd{V} = V/V_\mathrm{H}$, and $\nd{t} = t/T_\mathrm{int}$, is described by 
\begin{equation}
    \deriv{\mathcal{N}_V(\nd{V},\nd{t})}{\nd{t}}  = - \frac{\mathcal{N}_V(\nd{V},\nd{t})}{\avg{m}(\nd{V})} \int_0^{\nd{V}} \ndf(\nddelta;\nd{V}) \mathrm{d}{\nddelta} + \int_{\nd{V}}^\infty \mathcal{N}_V (\ndDelta,\nd{t}) \ndf(\nd{V};\ndDelta) \mathrm{d} \ndDelta, \label{eq:population_balance_abs}
\end{equation}
where the first term on the RHS gives the rate of consumption of bubbles of \dan{volume} $\nd{V}$ due their break-ups, and the second term on the RHS gives the rate of production of bubbles of \dan{volume} $\nd{V}$ due to the break-ups of larger bubbles \citep{Martinez-Bazan2010}. The break-up kernel $\nd{f}(\nddelta;\ndDelta) = f(\delta;\Delta) V_\mathrm{H} T_\mathrm{int}$ can be decomposed into a parent break-up frequency and \changed{volumetric child size distribution} with $\nd{f}(\nddelta;\ndDelta) = \nd{\omega}(\ndDelta) \nd{p}(\nddelta;\ndDelta)$, with the dimensionless break-up frequency $\nd{\omega}(\ndDelta) = \omega(d) T_\mathrm{int}$ and dimensionless \changed{volumetric child size distribution} $\nd{p}(\nddelta;\ndDelta) = p(\delta;\Delta) V_\mathrm{H}$. Thus, we can move $\nd{\omega}(\ndDelta)$ outside the integral in the first term on the RHS and invoke $\int_0^{\nd{V}} \nd{p}(\nddelta;\nd{V}) \mathrm{d} \nddelta = \avg{m}(\nd{V})$, \changed{with $\avg{m}(\nd{V})$ the average number of bubbles formed in the break-up of a bubble of volume $\nd{V}$,} to express the bubble consumption term as simply $-\mathcal{N}_V(\nd{V},\nd{t}) \nd{\omega}(\nd{V})$. \dan{Note that we define $\nd{p}(\nddelta;\ndDelta)$ so that it integrates over $\nddelta$ to the average number of child bubbles formed by the break-up of a bubble of volume $\ndDelta$.}

\subsection{Construction of the child size distributions}
\label{sec:Monte_Carlo}

We develop a parameterization of the break-up \changed{volumetric child size distribution} $\nd{p}(\nddelta;\ndDelta)$ that accounts both for child bubbles produced by both the slower inertial mechanism (occurring over the eddy turnover time) and the faster capillary pinching mechanism (occurring over the capillary timescale of the small child bubbles) \dan{using a Monte Carlo approach. We consider a set of rules constrained by our experimental and numerical observations describing the outcomes of individual break-up events, then aggregate the outcomes of these events into child size distributions.} 

\subsubsection{Statistics on the number of child bubbles formed}
\label{sec:model_m}

\begin{figure}
\centering
  \begin{overpic}[width=1\linewidth]{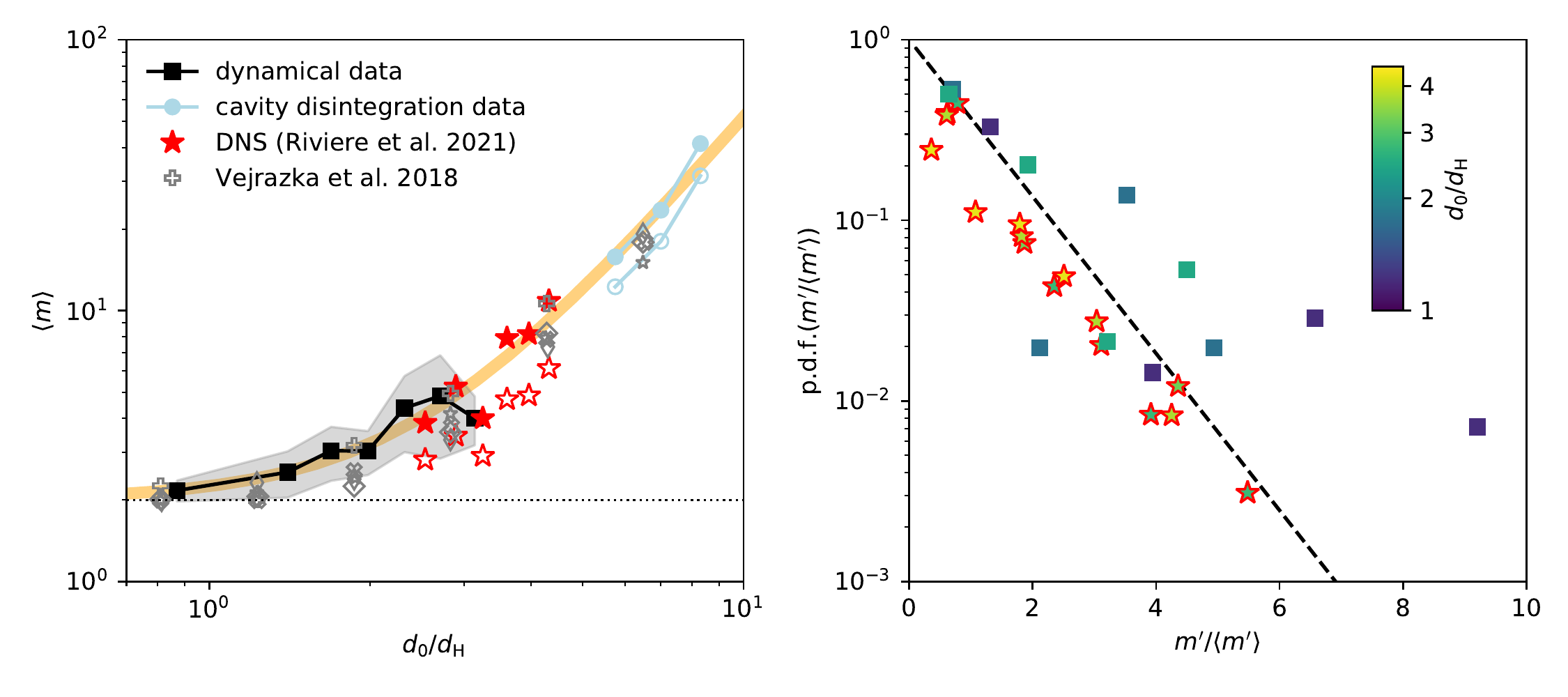}
  \put(1,42){(a)}
  \put(51,42){(b)}
  \end{overpic}
\caption{Experimental data on the number of child bubbles formed in each break-up. (a) The average number of resolved bubbles (with $d_\mathrm{min}/\dH = 0.07$) formed in each break-up event $\avg{m}$ as a function of the dimensionless parent bubble size. The shaded region shows $\pm$ one half of a standard around the mean for our dynamical data. Open circles give data from the disintegration of the three largest cavities, and closed circles give those data with an adjustment for the differing spatial resolution. Open stars give data from DNSs from \cite{Riviere2021jfm}, and closed stars give those data with the spatial resolution adjustment. The open gray markers give data from experiments reported by \cite{Vejrazka2018}. The thick orange line is the parameterization given in \cref{eq:avg_m_parameterization}. (b) The p.d.f. of \dan{$m'/\avg{m'}=(m-m_\mathrm{min})/(\avg{m} - m_\mathrm{min})$} for the experiments (squares) and DNSs (stars), along with the exponential fit employed in the Monte Carlo simulations. } 
\label{fig:dynamical_m_data}
\end{figure}

A key step in modeling each break-up is to constrain the \dan{distribution of the} number of bubbles formed in each event. \dan{To this end, we first} consider the data from our dynamical experiments given in \cref{sec:dynamical}. \changed{The average number of child bubbles larger than the experimentally-resolvable minimum size $d_\mathrm{min}/\dH \approx 0.07$, $\avg{m}$, is shown in \Cref{fig:dynamical_m_data} (a). As the parent bubble increases in size, more child bubbles are typically produced. Given the steep dependence of $\nd{p}(\nddelta;\ndDelta)$ on $\nddelta$, we must qualify each observation of $m$ with the minimum resolved bubble size to better enable comparisons between different experiments. For compactness, however, we take all $m$ values to be the number of resolved bubbles larger than $0.07 \dH$ unless otherwise noted.} Our experimental observations of $\avg{m}$, binned by $d_0/\dH$, are shown in the black squares, and the gray region around them bounds $\pm$ one half of a standard deviation around the mean.

Next, to consider the number of bubbles produced in the break-ups of larger bubbles, we turn to data from the disintegration of the three largest cavities presented in \cref{sec:air_cavity_results}. With the assumption that the initial splitting event happens nearly instantly after the bubble is released into the turbulence, to apply the same definition of the duration of the break-up, we define $\avg{m}$ for this dataset as the number of \changed{resolved} bubbles present after one eddy turnover time $T_\mathrm{turb}(d_0) = \epsilon^{-1/3} d_0^{2/3}$ has elapsed after the cavity release, which are denoted by the open circles in \cref{fig:num_vs_time_and_d0dH} (a) and \cref{fig:dynamical_m_data} (a). \changed{We invoke \cref{eq:number_adjustment} to apply a slight adjustment to these numbers in order to extrapolate results to the finer spatial resolution of the tracked break-up experiment, as discussed in \Cref{sec:number_adjustment}. The number of bubbles in the extrapolated range constitutes about $30\%$ of the ones in the observable range. These adjusted values are shown as the filled-in light blue circles in \Cref{fig:dynamical_m_data} (a).}

We have \dan{additionally} re-analyzed the DNSs of bubbles breaking in homogeneous, isotropic turbulence presented in \cite{Riviere2021jfm,Riviere2021cap}, tracking the bubble break-up events in a similar way to what has been done on the experimental data in \Cref{sec:dynamical}. From these DNSs, we can compute the average number of bubbles formed per event as a function of the parent bubble size, included in panel (a) as the red star markers. Open stars give the original observations, for which $d_\mathrm{min}/\dH = 0.25$, while the filled-in stars give the number adjusted for the spatial resolution. \dan{Note that while we consider $\mathrm{We}_\mathrm{c}=1$ for the experimental data, the value of $\dH$ for the DNS is given by $\mathrm{We}_\mathrm{c}=3$ \citep{Riviere2021jfm}.}

Finally, as a comparison, the open gray markers show the (un-adjusted) number of bubbles detected experimentally in break-ups by \cite{Vejrazka2018}, in which break-ups varied in $\epsilon$ and $d_0$ (which is denoted by the marker style). As shown in their paper, once collapsed to $d_0/\dH$, the dependence on the dimensional bubble size nearly disappears. 

The four datasets (our two experiments, those from \cite{Vejrazka2018}, and DNSs from \cite{Riviere2021jfm}) produce a coherent picture regarding the number of bubbles formed. When $d_0/\dH$ is small, break-ups tend to be binary, producing on average 2 child bubbles \dan{after $T_\mathrm{turb}(d_0)$}. As $d_0/\dH$ increases, the number of child bubbles increases. Surface tension is less effective at preventing the severe deformation of larger bubbles, leading to more complex deformed bubble shapes that yield a greater number of child bubbles. The orange curve in panel (a) shows a fit to the data of the form
\begin{equation}
    \avg{m} = m_\mathrm{min} + \frac{(\dodH)^{b_2}}{b_1}, \label{eq:avg_m_parameterization}
\end{equation}
where $m_\mathrm{min} = 2$ and the fit constants are $b_1 = 4$ and $b_2 = 2.3$.

\Cref{fig:dynamical_m_data} (b) compiles experimental and DNS data on the distribution of the number of child bubbles produced for increasing $\dodH$. The p.d.f.s of $m'/\avg{m'}$ are well-described by an exponential function $e^{-m'/\avg{m'}}$,   with $m'=m-m_\mathrm{min}$ and $\avg{m'} = \avg{m}-m_\mathrm{min}$, for both the experiments (shown as the squares) and DNS (shown as the stars).  Thus, for any parent bubble size we can write the p.d.f. of $m'$ as \changed{an exponential} distribution,
\begin{equation}
    r(m';\dodH) = \frac{\exp(-m' / \avg{m'})}{\avg{m'}}, \qquad m'>0, \label{eq:m_pdf}
\end{equation}
with $\avg{m'}+m_\mathrm{min}$ the mean number of children, a function of the parent bubble size. 

\subsubsection{A stochastic model for each break-up}
\label{sec:montecarlo_individual_breakup}

\begin{figure}
\centering
  \includegraphics[width=0.6\linewidth]{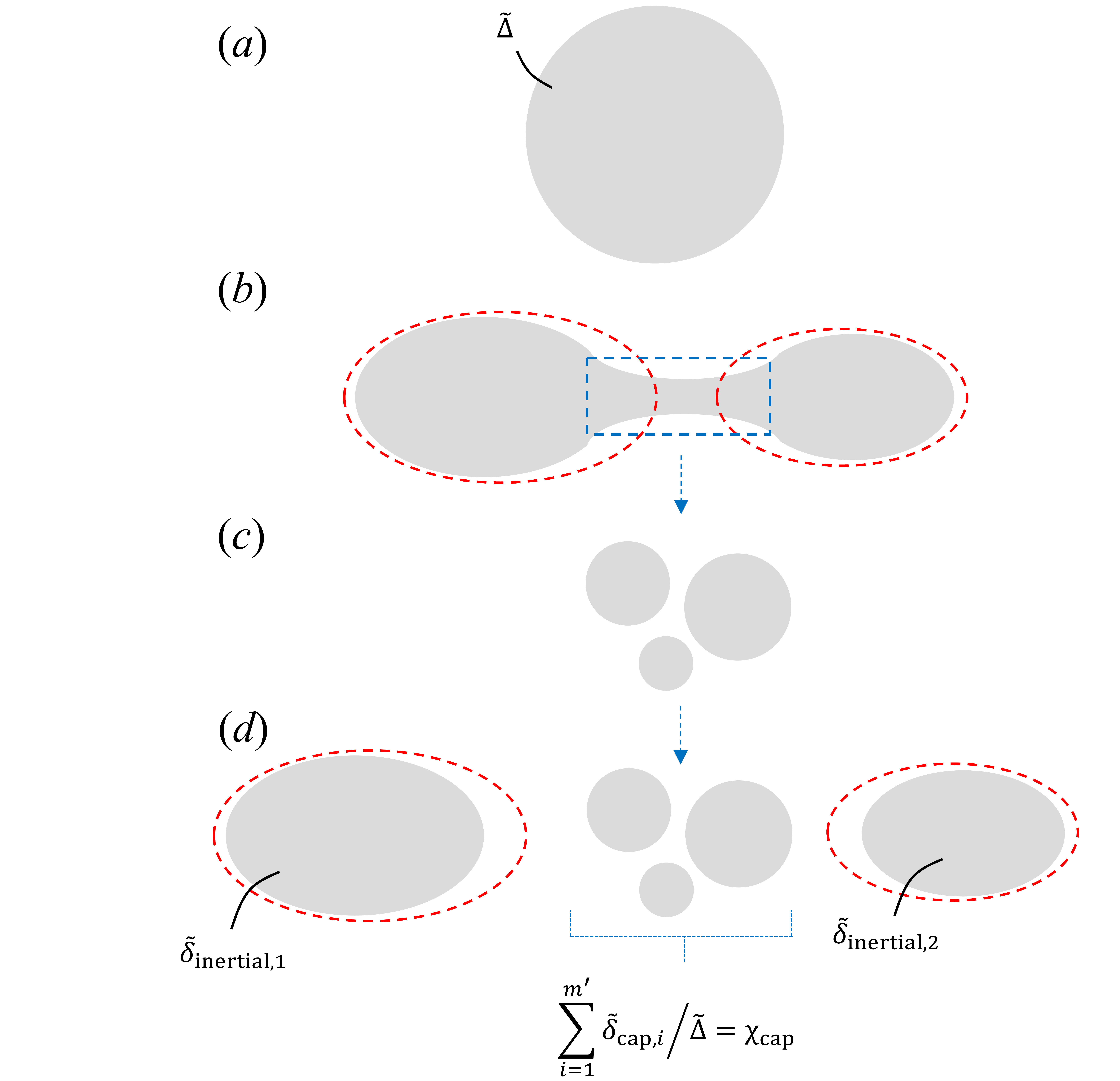}
\caption{Process of simulating one break-up for the Monte Carlo approach of a bubble of volume $\ndDelta$, shown in (a). (b) First, the bubble is taken to be deformed into two lobes, separated by a neck of gas. (c) Next, the sizes of the $m' = m-2$ capillary bubbles are picked from a $\delta_\mathrm{cap}^\alpha$ distribution. (d) Finally, the sizes of the two inertial bubbles $\nddelta_{\mathrm{inertial},i}$ are picked by from a uniform distribution over the remaining parent bubble volume (that which has not gone to the capillary bubbles).}
\label{fig:montecarlo_explanation_sketch_crop}
\end{figure}

The Monte Carlo approach involves running many iterations of a stochastic model and developing a statistical representation of the aggregated results. Each discrete simulation of a break-up mirrors the physical processes involved: the bubble, sketched in \Cref{fig:montecarlo_explanation_sketch_crop} (a), is first deformed into two lobes, shown in panel (b), and then some number of capillary bubbles are created as the neck separating the lobes collapses \dan{to create the} two inertial child bubbles.

For each iteration \dan{(i.e., one simulated breakup)} at a given value of $\ndDelta$, we first define the number of bubbles $m$ that will be produced by picking a value of $m'$ from the distribution $r(m';\dodH)$ given by \cref{eq:m_pdf}, adding $m_\mathrm{min} = 2$, and rounding to the nearest integer. We pick $\nddelta_\mathrm{min} = 0.07^3$ in order to match the experimental dataset on which the parameterization of $\avg{m}$ is based. As we will show, once the p.d.f.s have been constructed for this given value of $\nddelta_\mathrm{min}$, it will be straightforward to extend them to lower or higher values of $\nddelta_\mathrm{min}$.

For cases in which $m\geq 3$, the capillary mechanism produces $m' = m-2$ bubbles, whose sizes follow a $\propto \nddelta^{\alpha}$ distribution with $\alpha=-7/6$ \changed{(corresponding to the \dan{$\mathcal{P}_d(\ddH;\dodH) \propto (\ddH)^{-3/2}$} scaling described by \cite{Riviere2021cap}, as distributions in diameter are related to those in volume by $ \mathcal{P}_d(d/\dH;\dodH) = 3 (\ddH)^2 \nd{p}(\nddelta;\ndDelta)$ \citep{Martinez-Bazan2010,Qi2020})}. As is sketched in \Cref{fig:montecarlo_explanation_sketch_crop} (c), the \dan{volume} $\nddelta_{\mathrm{cap},i}$ of capillary bubble $i$ is picked from a power-law distribution with slope $\alpha$, bounded between $\nddelta_\mathrm{min}$ and the maximum allowable \dan{volume} for a capillary bubble given the previously-produced bubbles, $\nddelta_{\mathrm{cap,max},i}$. For the production of the first capillary bubble, we set $\nddelta_{\mathrm{cap,max},1} = \ndDelta$ (noting that the steep slope of \dan{$\mathcal{P}_d(\ddH;\dodH)$} with respect to $\ddH$ makes the production of capillary bubbles this large uncommon). For the production of the remaining capillary bubbles, we set $\nddelta_{\mathrm{cap,max},i} = \ndDelta - \sum_{j=1}^{i-1} \nddelta_{\mathrm{cap},j}$. At each step of the process, if $\nddelta_{\mathrm{cap},i}$ is greater than $\nddelta_{\mathrm{cap,max},i}/2$, we replace it with $\nddelta_{\mathrm{cap,max},i}/2 - \nddelta_{\mathrm{cap},i}$, such that for any splitting event, the smaller of the two produced does not further split.

Once the \dan{volumes} of the $m'$ capillary bubbles are specified, we must determine the \dan{volumes} of the two inertial bubbles. To that end, we first compute the portion of the parent bubble volume that has gone to the capillary bubbles, $\chi_\mathrm{cap} = \sum_{i=1}^{m'} \nddelta_{\mathrm{cap},i} / \ndDelta$. The size of the first of the two inertial child bubbles $\nddelta_{\mathrm{inertial},1}$ is drawn uniformly from the remaining bubble volume, $(1-\chi_\mathrm{cap}) \ndDelta$, and the second is taken as its complement, $\nddelta_{\mathrm{inertial},2} = (1-\chi_\mathrm{cap}) \ndDelta - \nddelta_{\mathrm{inertial},1}$. Once this is done, the volumes of all child bubbles produced in this single break-up have been determined.

\subsubsection{Aggregation of simulated break-ups into child size distributions}

For a given value of $\dodH$ (or the equivalent normalized volume \dan{$\ndDelta = (\dodH)^3$}), the process of simulating one break-up stochastically is repeated $n_\mathrm{MC} = 10^5$ times. 

\begin{figure}
\centering
  \begin{overpic}[width=1\linewidth]{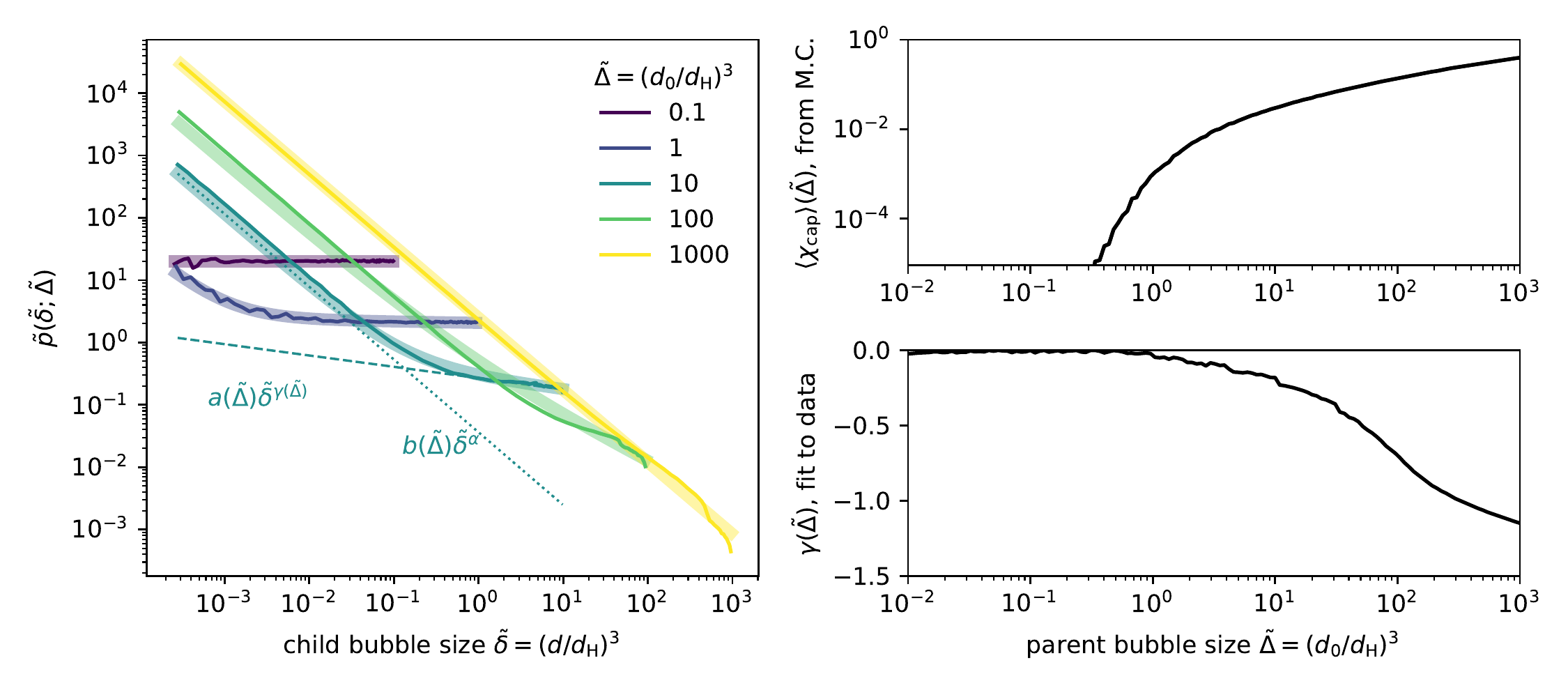}
  \put(14,39){(a)}
  \put(59,39){(b)}
  \put(59,19){(c)}
  \end{overpic}
\caption{\changed{Volumetric child size distributions} constructed via the Monte Carlo approach. (a) \changed{Volumetric child size distributions} $\nd{p}(\nddelta;\ndDelta)$ for five values of the parent bubble size $\ndDelta$. Distributions compiled from the Monte Carlo simulations are given by the thin lines, while the thick fainter lines give the fits using \cref{eq:child_size_dist_approx_twocomps}. The two components of the fit form of the distribution are illustrated for $\ndDelta = 10$. (b) The average capillary fraction $\avg{\chi_\mathrm{cap}}$ calculated from the ensemble of simulations, as a function of the parent bubble size. (c) Fit values of the exponent $\gamma(\ndDelta)$ employed in \cref{eq:child_size_dist_approx_twocomps}. Data for the curves in (b) and (c) and Python code to use them to construct $\nd{p}(\nddelta;\ndDelta)$ are \dan{will be made available online}.}
\label{fig:montecarlo_distributions_fit}
\end{figure}

For each $\ndDelta$, the sizes of the bubbles produced in each of the $n_\mathrm{MC}$ events are aggregated, \changed{and the distribution of all these child bubbles defines the \changed{volumetric child size distribution} $\nd{p}(\nddelta;\ndDelta)$. The distribution is normalized such that $\int_0^{\ndDelta} \nd{p}(\nddelta;\ndDelta) = \avg{m}(\ndDelta)$, with $\avg{m}(\ndDelta)$ the average number of bubbles formed.} Since the size distribution is aggregated from geometrically-plausible break-ups, it itself must satisfy any constraints relating to the sizes of the bubbles produced. \Cref{fig:montecarlo_distributions_fit} (a) shows the \changed{volumetric child size distributions} for five values of $\ndDelta$. When $\ndDelta$ is small, the child size distribution is nearly uniform, as the capillary production mechanism is negligible for small bubbles; for moderate $\ndDelta$, the child size distribution exhibits a $\nd{p} \propto \nddelta^\alpha$ scaling for small bubbles, while remaining close to flat for bubbles near the parent bubble size. For even larger bubbles, for which the capillary production mechanism is the most effective, the entire distribution approaches a $\nddelta^\alpha$ scaling.

For each $\ndDelta$, we also obtain $\avg{\chi_\mathrm{cap}}(\ndDelta)$, shown in \Cref{fig:montecarlo_distributions_fit} (b), by averaging the portion of the parent bubble volume going to the capillary child bubbles $\chi_\mathrm{cap}$ over the $n_\mathrm{MC}$ events. When $\ndDelta \ll 1$, $\chi_\mathrm{cap} \approx 0$, and essentially all of the parent bubble volume goes to the two inertial child bubbles. With larger $\ndDelta$, $\chi_\mathrm{cap}$ increases, reaching $\chi_\mathrm{cap} = 0.1$ at $\ndDelta = 60$. Even at $\ndDelta = 1000$, less than half of the parent bubble volume goes to the capillary bubbles.

We then fit each \changed{volumetric child size distribution} as a sum of two components, each stemming from one of the two mechanisms of child bubble production,
\begin{equation}
    \nd{p}(\nddelta;\ndDelta) = \underbrace{a(\ndDelta) \nddelta^{\gamma (\ndDelta)}}_\text{inertial mechanism}
    + \underbrace{b (\ndDelta) \nddelta^\alpha}_\text{capillary mechanism}, \label{eq:child_size_dist_approx_twocomps}
\end{equation}
with $\alpha=-7/6$ set by the distribution from which the capillary bubbles are picked and $\gamma(\ndDelta)$ chosen to match the aggregated Monte Carlo simulation data. The two remaining coefficients, $a(\ndDelta)$ and $b(\ndDelta)$, are constrained by the volume going to bubbles produced by each mechanism, leading to
\begin{align}
    a(\ndDelta) &=  (1 - \avg{\chi_\mathrm{cap}}) \left( \frac{ (\gamma + 2) \ndDelta }{\ndDelta^{\gamma+2} - \nddelta_\mathrm{min}^{\gamma+2}} \right),\\
    b(\ndDelta) &= \avg{\chi_\mathrm{cap}} \left( \frac{(\alpha+2)\ndDelta}{\ndDelta^{\alpha+2} - \nddelta_\mathrm{min}^{\alpha+2}} \right).
\end{align}
The fits to each child size distribution with \cref{eq:child_size_dist_approx_twocomps} are shown as the faint, thick lines in \Cref{fig:montecarlo_distributions_fit} (a).

\Cref{fig:montecarlo_distributions_fit} (c) shows the evolution of the exponent $\gamma(\ndDelta)$ describing the inertial production mechanism. Values of $\avg{\chi_\mathrm{cap}}(\ndDelta)$ and $\gamma(\ndDelta)$, which together contain all the necessary information about the child size distributions, are stored for many values of $\ndDelta$. To implement the child size distributions in a population balance model, we interpolate $\avg{\chi_\mathrm{cap}}(\ndDelta)$ and $\gamma(\ndDelta)$ for a given value of $\ndDelta$. Data for each curve and Python code to construct the child size distributions \dan{will be provided online at publication} for those wishing to implement the model we have constructed.


\subsection{Parameterization of the break-up frequency}

The next step is to parameterize how often the break-ups will occur. Here, using an approach that has been successfully applied to the break-up of oil droplets in turbulent jets \citep{Aiyer2019,Aiyer2020}, we integrate the effects of eddies smaller than the parent bubble size (each of \dan{dimensional} diameter $d_\mathrm{e}$) which contribute to break-up \citep{Prince1990,Tsouris1994}, yielding
\begin{equation}
    \nd{\omega}(\ndDelta) = K \frac{\dH}{\Lint} \int_0^{\dodH} \frac{\pi}{4} \left( \frac{d_0}{\dH} +  \frac{d_\mathrm{e}}{\dH} \right)^2 \nd{u}_\mathrm{turb}(d_\mathrm{e}/\dH) \left(\frac{d_\mathrm{e}}{\dH} \right)^{-4} \Omega(d_\mathrm{e}/\dH;d_0/\dH) \mathrm{d} (d_\mathrm{e}/\dH), \label{eq:Aiyer_breakuprate}
\end{equation}
where $K$ is an order-1 constant we adjust, $\nd{u}_\mathrm{turb}(d_\mathrm{e}/\dH) = C_2^{1/2} \epsilon^{1/3} d_\mathrm{e}^{1/3} / u' = C_\epsilon^{1/3} (d_\mathrm{e}/\dH)^{1/3} (\dH/\Lint)^{1/3}$ is the dimensionless turbulent velocity scale of the eddy, $(d_\mathrm{e}/\Lint)^{-4}$ is the approximate dimensionless eddy density \citep{Solsvik2016}, and $\Omega(d_\mathrm{e}/\dH;d_0/\dH)$ is the break-up efficiency given the eddy and bubble sizes. Neglecting viscous effects (given the low viscosity of air bubbles), the break-up efficiency, which gives the probability that an eddy has sufficient energy to overcome surface tension, is taken as the inverse of the exponential of the ratio between the \dan{average} change in surface energy associated with the break-up $E_\sigma(d_0)$ and the kinetic energy of the eddy $E_\mathrm{eddy}(d_\mathrm{e})$, \dan{$\exp(- E_\mathrm{eddy}(d_\mathrm{e}) / E_\sigma(d_0)$}. The \dan{average} surface energy change is given dimensionally by
\begin{equation}
    E_\sigma(d_0) = \frac{\sigma \pi}{4} \left( \int_{\delta_\mathrm{min}}^{\dan{\pi d_0^3/6}} p(\delta; \dan{\pi d_0^3/6}) \delta^{2/3} \mathrm{d} \delta - d_0^2 \right) = \Gamma \pi \sigma d_0^2 / 4,
\end{equation}
with the proportional change in surface area due to break-up $\Gamma$ dependent on the form of the child size distribution according to
\begin{equation}
    \Gamma(\ndDelta) = \frac{\int_{\nddelta_\mathrm{min}}^{\ndDelta} \nd{p}(\nddelta;\ndDelta) \nddelta^{2/3} \mathrm{d} \nddelta}{\ndDelta^{2/3}} - 1.
\end{equation}
The kinetic energy of the eddy is given by $E_\mathrm{eddy}(d_\mathrm{e}) = (\pi/4) \rho d_\mathrm{e}^3 C_2 (\epsilon d_\mathrm{e})^{2/3}$. Expressed in our non-dimensional units, the break-up efficiency is then
\begin{equation}
    \Omega(d_\mathrm{e}/\dH;d_0/\dH) = \exp \left( - \frac{ \Gamma(\ndDelta) (d_0/\dH)^2}{\mathrm{We}_\mathrm{c} (d_\mathrm{e}/d_\mathrm{H})^{11/3}} \right), \label{eq:breakup_efficiency}
\end{equation}
\dan{with the critical Weber number $\mathrm{We}_\mathrm{c}$ necessary to link the scales of the bubble and the turbulence.}

\begin{figure}
\centering
  \includegraphics[width=0.6\linewidth]{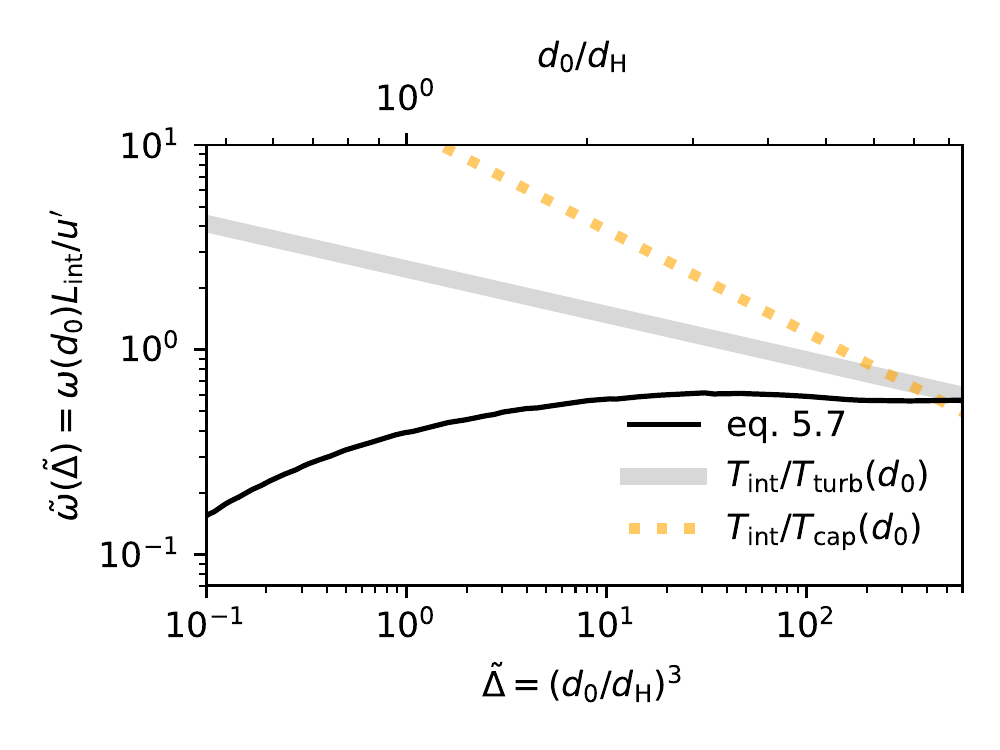}
\caption{The parent bubble break-up rate $\nd{\omega}$ as a function of its \dan{volume} $\ndDelta$, computed using the value of $\dH/\Lint$ for our dataset. The black line shows the parent bubble break-up frequency given by \cref{eq:Aiyer_breakuprate}. The thicker gray line gives the inverse of the eddy turn-over time at the parent bubble scale, which is taken to be the upper limit in the duration of each break-up event. The dotted orange line gives the inverse of the capillary timescale at the parent bubble scale.}
\label{fig:fitted_model_breakup_frequency}
\end{figure}

With each component specified, \cref{eq:Aiyer_breakuprate} is evaluated numerically and is shown in \Cref{fig:fitted_model_breakup_frequency}, using $K=2$ picked through a comparison to the experimental data given in \Cref{sec:air_cavity_results}. The break-up rate increases as bubbles approach the Hinze scale and then plateaus due to two competing effects: while larger bubbles are susceptible to a wider range of turbulent scales that may cause break-up, they tend to break into many more bubbles than smaller ones do, leading to a greater surface energy term in \cref{eq:Aiyer_breakuprate}. This means that while more eddies are interacting with the parent bubble, each is less likely to have sufficient energy to cause a break-up.

The thicker gray line in \Cref{fig:fitted_model_breakup_frequency} gives the inverse of the turbulent turn-over time at the parent bubble scale, which we take to set the duration of each break-up event. The break-up frequency is thus consistent with the break-up duration, since $\nd{\omega}(\ndDelta) = \dan{g} T_\mathrm{int}$ being strictly less than $T_\mathrm{int} / T_\mathrm{turb}(d_0)$ means that the typical duration of a break-up is never longer than the typical time between such \dan{break-ups. Finally,} the dotted orange line gives the inverse of the (dimensionless) capillary timescale at the parent bubble scale, $T_\mathrm{int}/T_\mathrm{cap}(d_0)$, showing that capillary effects happen faster than both the break-up duration and time between break-ups (up until the largest bubbles we consider). The capillary pinching events responsible for sub-Hinze bubble creation thus occur over even shorter durations, as the capillary timescales of the small child bubbles formed will be much faster than that of the parent bubble.

\subsection{Summary of parameters involved in the model}

\changed{To summarize, \Cref{tab:model_parameters} lists each parameter in the model and explains how each is determined.

\begin{table}
\changed{
\begin{tabularx}{\linewidth}{p{1.5cm} l | p{2cm} XX}
\multicolumn{1}{c}{Model element}            &  \multicolumn{1}{c}{Equation} & \multicolumn{1}{c}{Variable} & \multicolumn{1}{c}{Description}  & \multicolumn{1}{c}{Constraints} \\
\hline
number of bubbles produced & \cref{eq:avg_m_parameterization} & $\Delta T_\mathrm{break-up} = T_\mathrm{turb}(d_0)$   & break-up duration (over which child bubbles are formed) & theory (\Cref{sec:model_timescales}), informed by experimental and numerical data \citep{Risso1998,Riviere2021jfm} \\
&  & $b_1 = 4$  & prefactor for number of bubbles formed per break-up & fit to our experimental and numerical data (\Cref{fig:dynamical_m_data})  \\
 &  & $b_2 = 2.3$ & power-law exponent in parent volume for number of bubbles &  \\
child size distribution shape & \cref{eq:child_size_dist_approx_twocomps} & $\alpha = -7/6$  & power-law exponent for the capillary contribution, corresponding to $N(d) \propto d^{-3/2}$  & theory \citep{Riviere2021cap} \\ 
 & & $a(\ndDelta)$ & magnitude of the capillary contribution  & Monte Carlo simulation results (\Cref{fig:montecarlo_distributions_fit})  \\ 
 &  & $b(\ndDelta)$  &  magnitude of the inertial contribution &  \\
 &  & $\gamma(\ndDelta)$  & power-law exponent for the inertial contribution  &   \\ 
break-up frequency & \cref{eq:Aiyer_breakuprate}  & $K = 2$  & break-up frequency prefactor& fit to transient experimental data, within the range suggested by \cite{Aiyer2019} \\
 & &  $C_2 = 2.0$ & $D_\mathrm{LL}(d) / (\epsilon d)^{2/3}$ in inertial subrange for HIT          & \cite{Pope2000}                              \\
 & &  $C_\epsilon = 0.7$  & $\epsilon L_\mathrm{int} / u'^3$ for HIT                                 & \cite{Sreenivasan1997}                    \\
  & \cref{eq:breakup_efficiency} &  $\mathrm{We}_\mathrm{c} = 1$ & critical Weber number & experimental break-up threshold
\end{tabularx}
\caption{Parameters involved in the bubble break-up model, their physical origin and the experimental/numerical data constraints.}
\label{tab:model_parameters}}
\end{table}
}


\subsection{Model comparison to transient air cavity disintegration data}

\begin{figure}
\centering
  \begin{overpic}[width=1\linewidth]{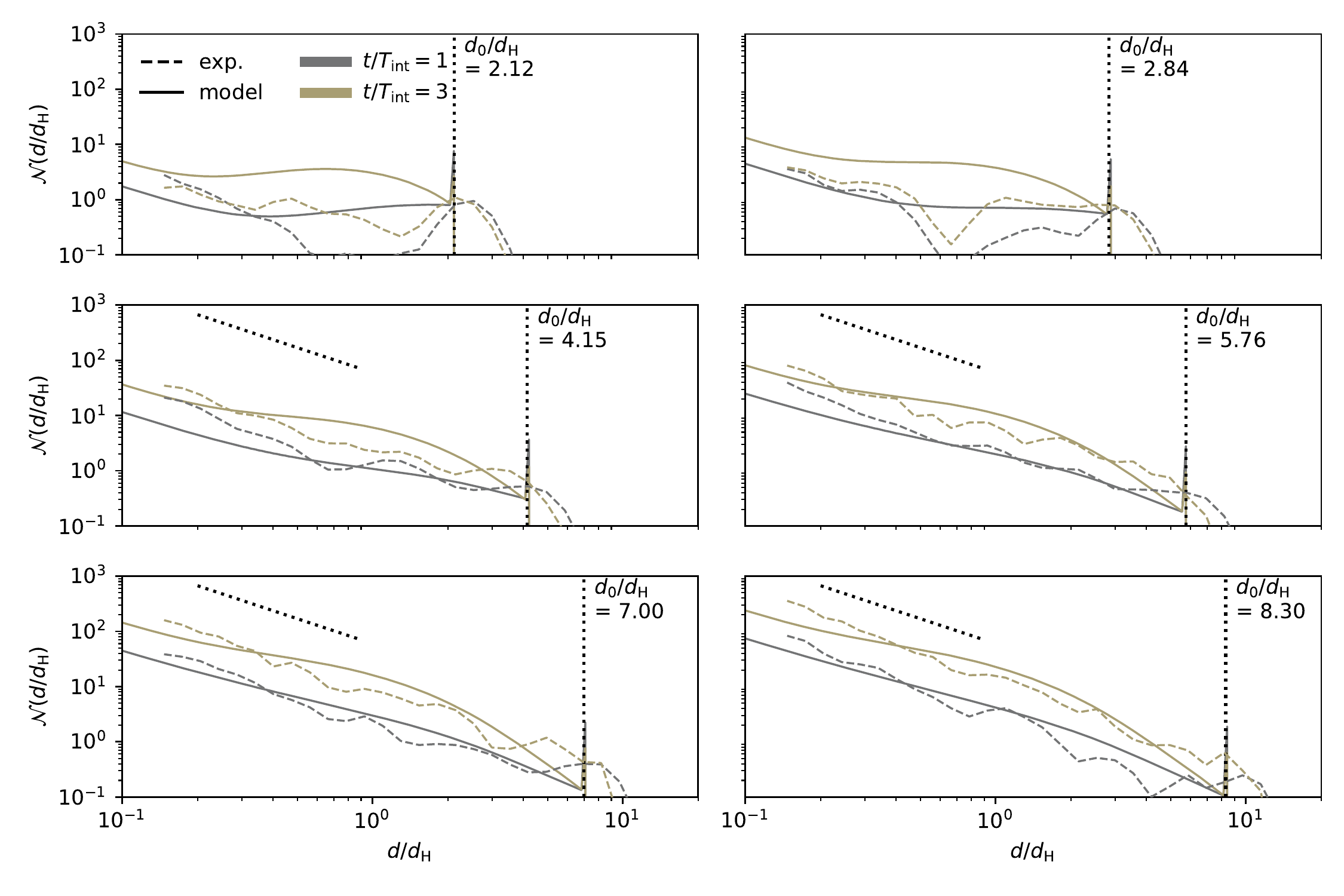}
  \put(2,61){(a)}
  \put(56,61){(b)}
  \put(2,41){(c)}
  \put(56,41){(d)}
  \put(2,21){(e)}
  \put(56,21){(f)}
  \end{overpic}
\caption{Comparisons of the experimental and modeled values of $\mathcal{N}(\ddH)$ at $t/T_\mathrm{int} = 1$ and $3$ for each value of $\dodH$. The dotted vertical line gives the value of $\dodH$ for each condition. Dotted lines give the $\mathcal{N}(\ddH) \propto (\ddH)^{ -3/2}$ sub-Hinze scaling. Good agreement between the measured and modeled distributions are observed for the full range of $d_0/d_H$ and times.}
\label{fig:fitted_model_comparisons}
\end{figure}

With $\nd{p}(\nddelta;\ndDelta)$ and $\nd{\omega}(\ndDelta)$ now fully specifying $\nd{f}(\nddelta;\ndDelta)$, we can simulate the turbulent disintegration of cavities we studied experimentally in \Cref{sec:air_cavity_results} by picking the appropriate initial condition for each (i.e., $\mathcal{N}(\ddH)$ giving one bubble of size $\dodH$) and integrating \cref{eq:population_balance_abs} in time. \Cref{fig:fitted_model_comparisons} compares the experimental and modeled vales of the dimensionless bubble \changed{size} distribution $\mathcal{N}(\ddH)$ at $t/T_\mathrm{int} = 1$ and 3 for each value of $\dodH$, with $\dodH=2.1$ in panel (a) and $\dodH=8.3$ in panel (f).


First, the model accurately reproduces the observed magnitudes of the size distributions near the Hinze scale, both in time and in the initial cavity size. Second, an $\mathcal{N}(\ddH) \propto (\ddH)^{-3/2}$ scaling is approached for $\ddH < 1$ with larger $\dodH$, and this scaling is adopted more rapidly with larger cavities. With $\dodH = 2.1$ and 2.84, shown in panels (a) and (b), $\mathcal{N}(\ddH)$ is flat near the Hinze scale at $t/T_\mathrm{int} = 1$, as the child size distributions for parent bubbles of these cavity sizes are largely flat (as shown in \Cref{fig:montecarlo_distributions_fit} (a)). With larger parent cavities, the sub-Hinze distribution steepens as the capillary mechanism contributes more significantly to the child size distributions for parent bubbles of these larger cavity sizes (as evidenced in the $\avg{\chi_\mathrm{cap}}(\ndDelta)$ curve shown in \Cref{fig:montecarlo_distributions_fit} (b)).


\section{Conclusions}
\label{sec:breakup_conclusions}

\changed{In this paper, we used results from two sets of experimental measurements to describe the production of bubbles smaller than the Hinze scale by turbulent bubble break-up. We experimentally demonstrate that a $N(d) \propto d^{-3/2}$ scaling for bubbles smaller than the Hinze scale ($d < \dH$) is obtained with the break-up of air cavities much larger than the Hinze scale subjected to forced turbulence, experimentally studying cavities up to $d_0 = 8.3 \dH$ with accurate measurements of bubble sizes down to approximately $0.1 \dH$. The $N(d)$ scaling we find is similar to the one reported in measurements and simulations of bubble size distributions under breaking waves \citep{Deane2002,Wang2016,Mostert2021}.

The small bubbles that are produced are significantly separated in size from the turbulent motions which are strong enough to cause break-up. Thus, the link between their sizes and the turbulent motions which do instigate break-up necessarily involves additional physics. Following \cite{Riviere2021cap}, we identify the capillary instability of deformed bubble ligaments which are involved in larger-scale turbulent deformations as the mechanism responsible for small bubble production. Crucially, significant small bubble production by this mechanism is limited to parent bubbles with $d_0 \gg \dH$, as only bubbles much larger than the Hinze scale can become deformed to a severe enough extent to produce the ligaments from which the small bubbles originate.

The first piece of evidence we provide for this role of capillarity is visual: \Cref{fig:explanation_figure_small_fig,fig:instability_cases} show a number of instances of small bubbles being left behind after the collapse of gas ligaments. Second, the experimental $N(d) \propto d^{-3/2}$ scaling for $d < \dH$ with $d_0 \gg \dH$ is coherent with the $P(d) \propto d^{-3/2}$ scaling for the break-up child size distribution reported by \cite{Riviere2021cap}, who showed that the lifetime of ligaments before their collapse to produce a bubble of size $d$ coincides with the capillary time scale of a bubble of size $d$, $T_\mathrm{cap} \propto d^{-3/2}$. 

We implemented these physical ideas in a population balance model of turbulent bubble break-up. The child size distributions describing individual break-up events were constructed with a Monte Carlo approach involving simulations of many break-ups. The statistics of each simulated break-up are prescribed by our understanding of the role of capillarity and additional experimental results on individual bubble break-up in which parent and child bubbles were tracked dynamically in three dimensions. The resulting expression for the child size distribution, \cref{eq:child_size_dist_approx_twocomps}, involves two components: one describes the effect of the large-scale deformation to a parent bubble by an energetic turbulent eddy, and the other describes the action of capillarity in producing small bubbles. Finally, the rate at which parent bubbles undergo break-ups was determined by integrating the action of eddies below the bubble's size, which all contribute to break-up. The complete model (consisting of the child size distributions and the parent break-up frequency) yields a good match to our transient experimental data.}

Along with the recent analysis of DNSs of bubble break-up in turbulence from \cite{Riviere2021jfm,Riviere2021cap}, this experimental work opens the door to a new understanding role of capillarity in turbulent bubble break-up, in which surface tension not only counteracts the initial turbulent deformation to a bubble but also leads to the formation of sub-Hinze bubbles through capillary instabilities that arise during the final stages of the break-up process.

\section*{Acknowledgements}
\label{sec:acknowledgements}
This work was supported by the NSF CAREER award 1844932 to L.D.

We declare no conflict of interest.

\appendix
\section{Appendix: air cavity disintegration data processing}

\subsection{Identification of bubble sizes}
\label{sec:cavity_bubble_detection}

Bubble sizes are detected with image processing of the images of the cavity disintegrations in multiple stages. First, a simple image intensity threshold is applied to binarize each greyscale image, and the bright spots at the interior of each bubble image is filled in. A first pass at extracting the bubble diameter $d_\mathrm{intensity}$ based on this intensity threshold is then made by computing the equivalent diameter of a circle with the same projected area as the binarized bubble image. As we find that the determined sizes of small bubbles ($d_\mathrm{intensity} < d_\mathrm{cutoff}$, with $d_\mathrm{cutoff} = \SI{1.5}{mm}$) are sensitive to the image intensity threshold chosen, we individually employ a Canny filter \citep{Canny1986} to the images of each of these small bubbles to find their borders. Their diameter $d$ is then defined as the equivalent diameter of the projected area inside the bubble border. For larger bubbles (with $d_\mathrm{intensity} > d_\mathrm{cutoff}$, for which the Canny edge detection often fails due to the deformed bubble shape), we define the diameter as $d = d_\mathrm{intensity} + \sigma_d$, where $\sigma_d = \SI{-25}{\micro m}$ is the typical value to which $d_\mathrm{Canny} - d_\mathrm{intensity}$ asymptotes for bubbles approaching $d_\mathrm{cutoff}$.

\subsection{Adjusting the size distribution to account for bubble advection and smoothing of the size distributions}
\label{sec:advection_correction}

The time-dependent size distribution $\mathcal{N}(\ddH,t/T_\mathrm{int})$ is first computed on a frame-by-frame basis, and then is averaged in time over a window with width $\tau/T_\mathrm{int}$ that increases with $t/T_\mathrm{int}$. The width $\tau$ is first set as $\tau/T_\mathrm{int} = 0.13 + 0.38 (t/T_\mathrm{int})$ and then clipped at $\tau/T_\mathrm{int} = 2.67$, which it reaches at $t/T_\mathrm{int} = 6.68$.

\begin{figure}
\centering
  \includegraphics[width=0.9\linewidth]{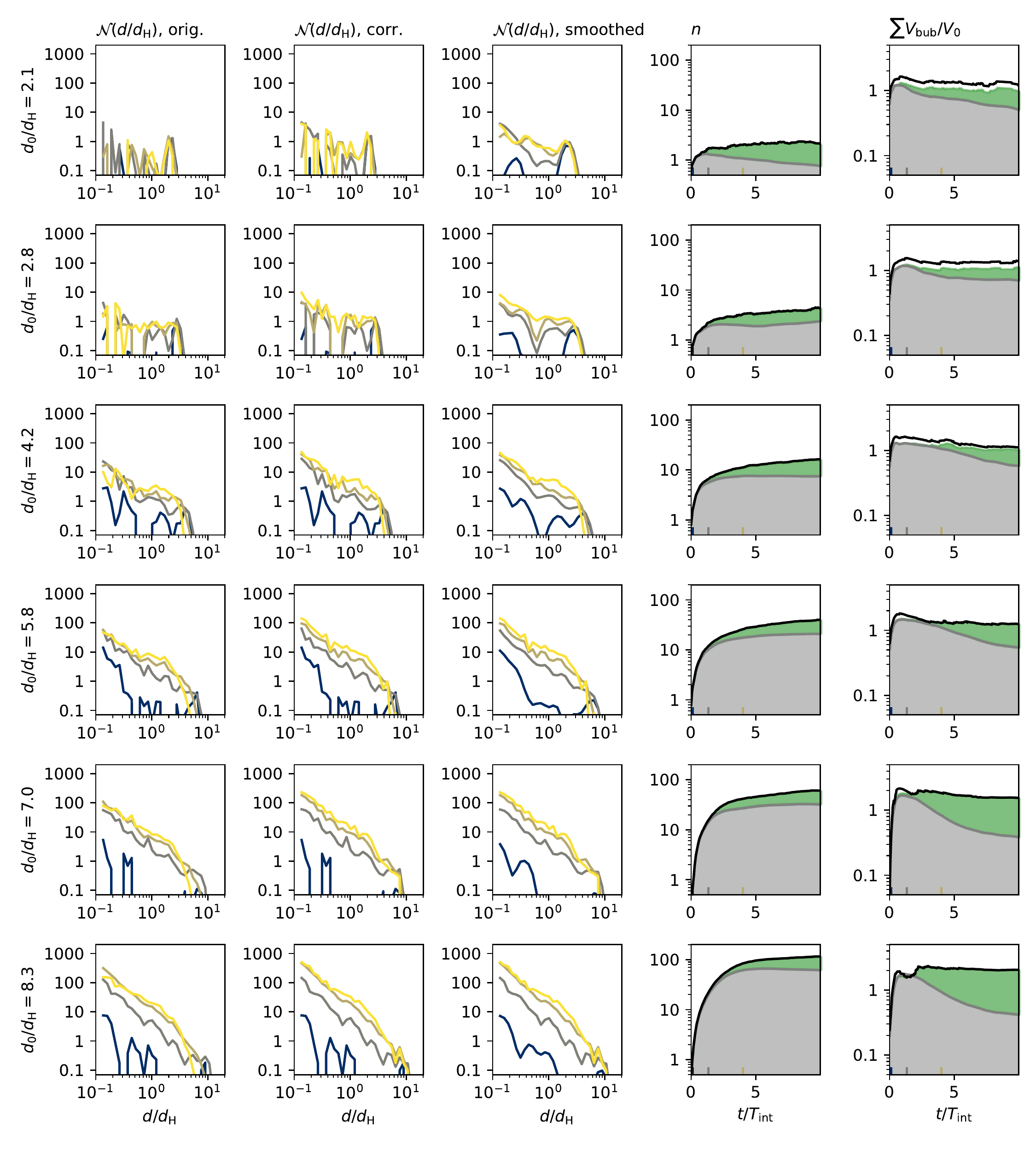}
\caption{Visualization of the adjustment to the bubble size distributions based on the advection of bubbles out of and into the field of view for each size air cavity studied. (a) The original (unadjusted) size distributions of the bubbles in-frame at four times. (b) The adjusted size distributions. (c) The adjusted size distributions with the slight smoothing applied, which we consider in the paper. (d) The total number of bubbles detected in-frame (gray) and the total number of bubbles, including the advection adjustment (black).  (e) The sum of the volumes of the bubbles detected over time, normalized by the known volume of the air cavity. The gray curve is the volume of bubbles detected in-frame; the black curve is that curve added to the volume of bubbles from the advection adjustment, in green.}
\label{fig:advection_correction_explanation}
\end{figure}

Due to the buoyant rise of the bubbles and their advection by the turbulence in the air cavity disintegration experiments, bubbles leave the measurement region over time. This complicates the analysis of our data: if we were to solely consider bubbles viewed in-frame, we would calculate a rapid loss of bubble volume as bubbles leave the field of view. Results would be further skewed by any size dependence of the bubbles' motions. Indeed, the transient bubble size distributions based on the bubbles viewed in-frame shown in the left of \Cref{fig:advection_correction_explanation} exhibit a nonphysical decrease in $\mathcal{N}(d/d_\mathrm{H})$ for small $d$ at later times. Similarly, the total volume and number of bubbles tracked, shown in the fourth and fifth columns respectively as the gray lines, decreases as bubbles leave the field of view. %

We address this experimental limitation by tracking the bubbles' motion in two dimensions and making note of when bubbles leave or enter the measurement region near one of its four borders. Then, to compute $N(d)$ for some time $t$, along with the bubbles of size $d$ imaged at time $t$, we add counts of all the bubbles of size $d$ that have left the measurement region before time $t$, and subtract counts of all the bubbles of size $d$ that have entered the measurement region before time $t$. The resulting adjusted size distributions are shown in the second column of \Cref{fig:advection_correction_explanation}.

The final step in the processing is to slightly smooth each $\mathcal{N}(\ddH,t)$ curve in bubble size by an amount dependent on the total number of bubbles present at each time, $n(t) = \int_0^\infty \mathcal{N} \mathrm{d}(\ddH)$. Size distributions are computed with 30 geometrically-spaced bins between $\ddH = 0.14$ and 17.5. Then, at each time, each curve is smoothed with a Gaussian filter with a standard deviation of $\sigma_\mathrm{smooth}(t)$ bins, with $\sigma_\mathrm{smooth}(t)$ picked given an empirical function of the number of bubbles present. When fewer than three bubbles are present, $\sigma_\mathrm{smooth}(t)$ is set to one bin; when more than thirty are present, $\sigma_\mathrm{smooth}(t)$ is 0.1 bins. $\sigma_\mathrm{smooth}(t)$ is interpolated between these two limits when a moderate number of bubbles are present. This approach ensures minimal smoothing when a sufficient number of bubbles are present and a moderate amount of smoothing when few bubbles are present. The smoothed size distributions, which we consider in the paper, are shown in the third column of \Cref{fig:advection_correction_explanation}.

This advection adjustment approach effectively "freezes" in place the record of bubbles as they leave the measurement volume. The green regions in the third and fourth columns of \Cref{fig:advection_correction_explanation} show the additional number of bubbles and corresponding additional bubble volume added to the bubble record with this method. The black lines show the sum of the in-frame measurements and this adjustment. Any limit obtained by the adjusted $n$ curve is still not especially physically meaningful, as some of the bubbles which have exited the field of view are not much smaller than the Hinze scale, so they would eventually break apart further if left within the turbulence region. However, it is qualitatively closer representation of the "true" behavior than would be obtained through just the in-frame measurements.

The plots of the summed bubble volume (normalized by the cavity volume) shown in the fourth column of \Cref{fig:advection_correction_explanation} reveal a second limitation in our bubble detection method: the total volume of bubbles considered, $\int_0^\infty V N_V(V) \mathrm{d} V$, is not a constant value equal to the known volume of the air cavity $V_0$. At early times, the total bubble volume is under-counted, and becomes over-counted at later times. This is due to the highly-deformed shapes of large bubbles, for which the \changed{equivalent diameter determination we employ is only a rough approximation}. Further, bubbles whose images overlap can be detected as a single larger bubble.

We note that, aside from some representative images taken at late times, the latest measurement of $\mathcal{N}(d/d_\mathrm{H})$ presented in the paper or employed in our analysis is $t/T_\mathrm{int}=5$, at which point the advection adjustment has had only a moderate impact on the size distribution for all cavity sizes.

\section{Comparison of cavity disintegration with and without turbulence}
\label{sec:quiescent_breakup_comparison}

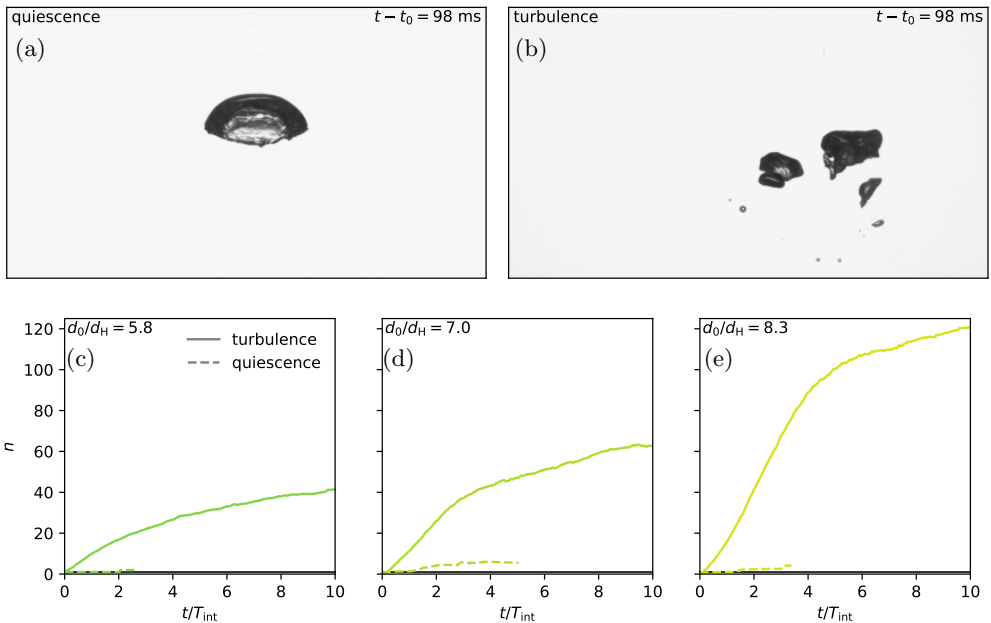
\begin{figure}
\centering
  \begin{overpic}[width=1\linewidth]{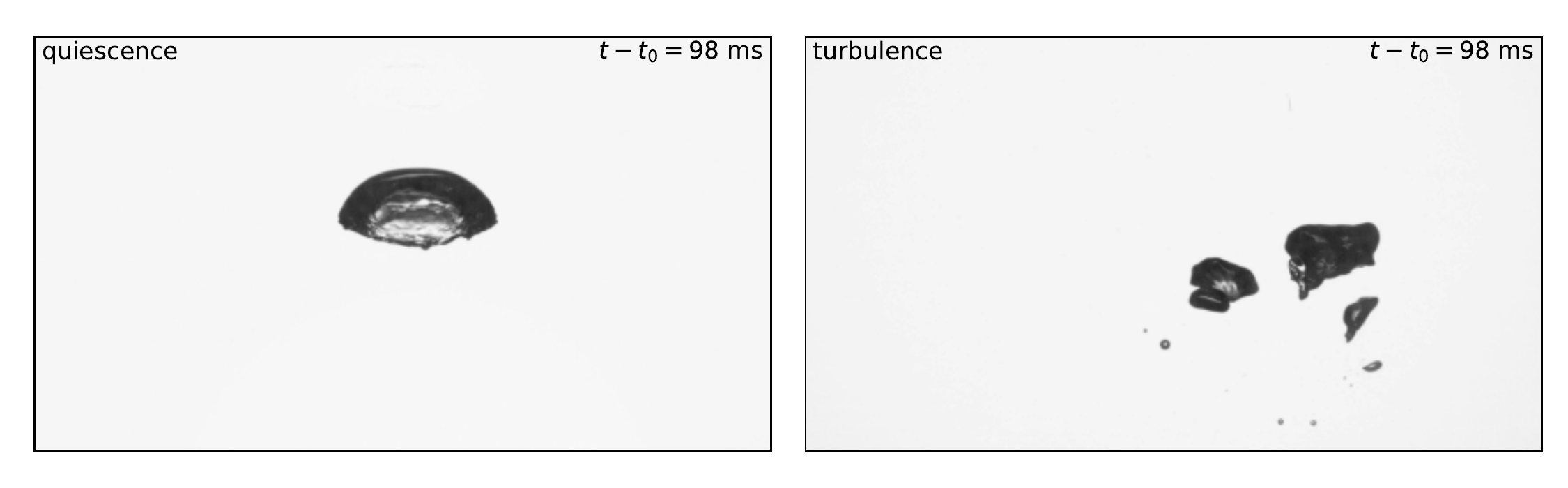}
  \put(3,24){(a)}
  \put(52,24){(b)}
  \end{overpic}\\
  \begin{overpic}[width=1\linewidth]{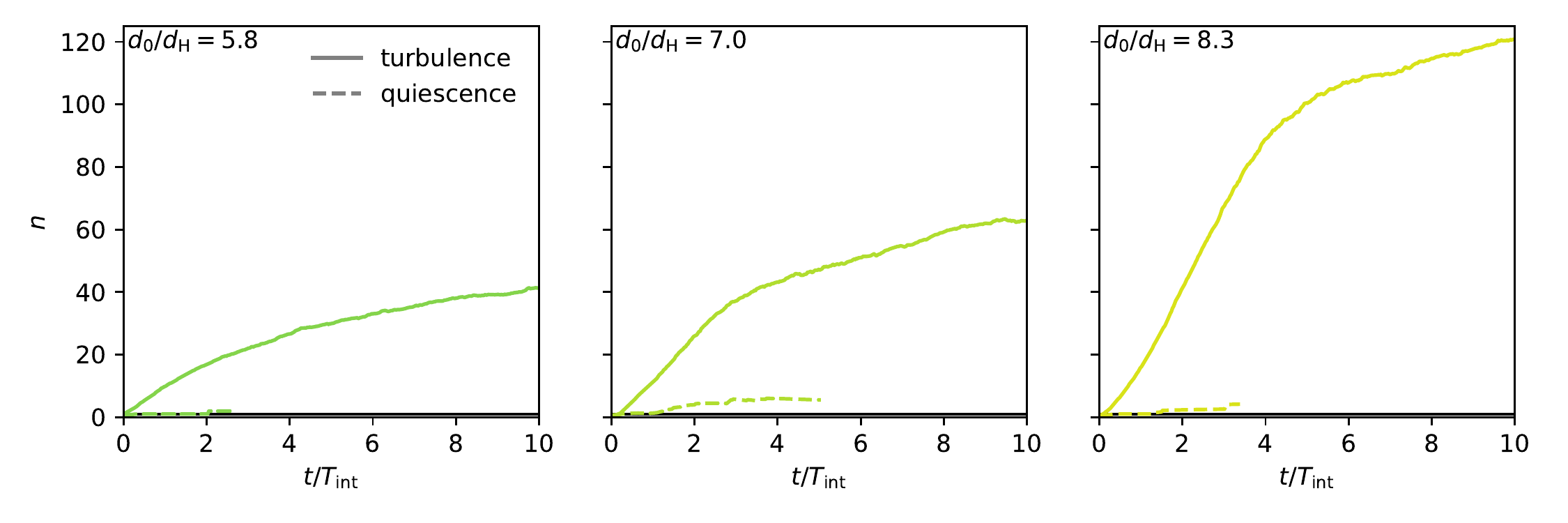}
  \put(8,27){(c)}
  \put(39,27){(d)}
  \put(70.5,27){(e)}
  \end{overpic}
\caption{\changed{The cavity release experiments with and without turbulence. (a-b) A comparison between images taken in quiescence (a) and turbulence (b) for a bubble with $d_0/d_\mathrm{H} = 5.8$ (when in turbulence). (c-e) A comparison of the transient number of bubbles present for initial cavities of varying sizes, in turbulence (solid lines) and in quiescence (dashed lines). The quiescent cases considered are limited to those within the range of cup spin velocities taken in the experiments with turbulence.}}
\label{fig:n_vs_time_turb_vs_quiescence}
\end{figure}

\changed{The cavity release experiment detailed in \Cref{sec:exp_cavity} was also run with the turbulence-generating pumps turned off, such that the cavity was released into otherwise still water. In these experiments, the extent of the bubble production is greatly reduced. \Cref{fig:n_vs_time_turb_vs_quiescence} (a) and (b) shows snapshots of two bubbles of the same size, at the same time after their release, into quiescence and turbulence, respectively. The bubble released into quiescence is deformed by buoyancy but has not broken apart; the bubble released by turbulence has undergone break-ups. Bubbles released into quiescence occaisionally break due to the cup's motion or their large buoyant deformations \citep{Landel2008}, but \Cref{fig:n_vs_time_turb_vs_quiescence} (c-e), which show the number of bubbles produced over time in turbulence and quiescence for the three largest cavity sizes, indicate that the moderate bubble production from these break-ups is negligible compared to the much greater production in turbulence.}

\section{Adjustment of the number of resolved bubbles to account for the minimum resolved child size}
\label{sec:number_adjustment}

\changed{Since we find a bubble size distribution that scales as $N(d) \propto d^{\alpha_d}$ with $\alpha_d < -1$ for small bubbles, the total number of bubbles above some minimum size will diverge as that minimum size decreases, up until some additional physical limit is encountered. Therefore, to enable a more direct comparison between datasets in which the experimental or numerical resolution differs, we can adjust the total number of bubbles formed in a break-up to account for the different resolved sizes.

Let us denote by $m[d> z \dH]$ the average number of resolved bubbles larger than $z \dH$ that are formed in a break-up. Given a known value of $m[d> x \dH]$, we can find the corresponding value of $m[d > y \dH]$, which is the hypothetical number resolved had a minimum spatial resolution of $y \dH$ been employed. Following the conceptual model presented in \cref{sec:concurrent_mechanisms}, we assume that all but two of the child bubbles produced in each break-up follow a power-law scaling $\propto (d/\dH)^{\alpha_d}$, with $\alpha_d=-3/2$. With this assumption, we calculate the appropriate prefactor for the sub-Hinze distribution given the observed value of $m[d> x \dH]$, then extend the distribution to $d_\mathrm{min}/\dH = y$ and integrate over all the larger bubble sizes to get the effective number in the range that is resolvable in the hypothetical experiment, yielding 
\begin{equation}
    m[d>y d_\mathrm{H}] = \left( m[d>x d_\mathrm{H}]-2 \right) \left( \frac{(\dodH)^{(\alpha_d)+1} - y^{(\alpha_d)+1}}{(\dodH)^{(\alpha_d)+1} - x^{(\alpha_d)+1}} \right) + 2. \label{eq:number_adjustment}
\end{equation}
}

\bibliography{references}

\end{document}